%% file: mainardi_BUDAPEST97.tex
\documentclass[11pt]{article}
\usepackage[dvips]{graphicx}
\begin{document}
\tolerance=10000


\hyphenpenalty=2000
\hyphenation{visco-elastic visco-elasticity}
\setcounter{page}{1}
\thispagestyle{empty}




\font\note=cmr10 at 10 truept  
\newcommand{\eproof}{\rule{0.2cm}{0.2cm}}

\newcommand{\stt}{\small\tt}

\tolerance=10000
\def\pni{\par \noindent}
\def\vsh{\smallskip}
\def\vs{\medskip\noindent}
\def\vvs{\bigskip\noindent}
\def\vvvs{\bigskip\medskip\noinden} 
\def\vsp{\vsh\pni}
\def\vsn{\vsh\pni}
\def\cen{\centerline}
\def\ra{\item{a)\ }} \def\rb{\item{b)\ }}   \def\rc{\item{c)\ }}
\def\eg{{\it e.g.}\ } \def\ie{{\it i.e.}\ }
\def\rstar{\item{$\null ^*$)\ }}
   \def\rl{\item{-- \ }}
    \def\rp{\item{}}
    \def\rf#1{\item{{#1}.\ }}  
\def\Prob{\hbox{{\rm Prob}}}
\def\e{\hbox{{\rm e}}}
\def\exp{\hbox{{\rm exp}}}
\def\tan{\hbox{{\rm tan}}}
\def\atan{\hbox{{\rm atan}}}
\def\arctan{\hbox{{\rm arctan}}}
\def\log{\hbox{{\rm log}}}
\def\ln{\hbox{{\rm ln}}}
\def\with{\hbox{{\rm with}}}
\def\sign{\hbox{{\rm sign}}}
\def\Ai{\hbox{{\rm Ai}}}
\def\erf{\hbox{{\rm erf}}}
\def\erfc{\hbox{\rm erfc}}
\def\ds{\displaystyle}
\def\dis{\displaystyle}
\def\q{\quad}	 \def\qq{\qquad}
\def\lan{\langle}\def\ran{\rangle}
\def\l{\left} \def\r{\right}
\def\lra{\Longleftrightarrow}
\def\d{\partial}
\def\dr{\partial r}  \def\dt{\partial t}
\def\dx{\partial x}   \def\dy{\partial y}  \def\dz{\partial z}
\def\rec#1{{1\over{#1}}}
\def\bar{\tilde}
\def\barr{\widetilde}
\def\hatt{\widehat}
\def\epsilons{{\widetilde \epsilon(s)}}
\def\sigmas{{\widetilde \sigma (s)}}
\def\fs{{\widetilde f(s)}}
\def\Js{{\widetilde J(s)}}
\def\Gs{{\widetilde G(s)}}
\def\Fs{{\wiidetilde F(s)}}
 \def\Ls{{\widetilde L(s)}}
\def\L{{\cal L}} 
\def\F{{\cal F}} 
\def\NN{\hbox{\bf N}}
\def\RR{\hbox{\bf R}}
\def\II{\hbox{\bf I}}
\def\CC{\hbox{\bf C}}
\def\ZZ{{\bf Z}} 
\def\D{{\cal D}}  
\def\Gc{{\cal {G}}_c}	\def\Gcs{\barr{\Gc}} 
\def\Gs{{\cal {G}}_s}	\def\Gss{\barr{\Gs}} 
\def\args{(x/ \sqrt{\D})\, s^{1/2}}
\def\argsa{(x/ \sqrt{\D})\, s^{\beta}}
\def\arg{ x^2/ (4\,\D\, t)}



\font\title=cmbx12 scaled\magstep2
\font\bfs=cmbx12 scaled\magstep1

 \cen{FRACALMO PRE-PRINT   {\bf www.fracalmo.org}}

\vs

\hrule

\vsh
\begin{center}

{\title Probability distributions generated by}
\vs

{\title fractional diffusion equations}\footnote{
This paper is based on an invited talk given
by Francesco Mainardi
at the International {\it Workshop on Econophysics} held
at Bolyai College, E\"otv\"os University, Budapest,
on July 21-27, 1997.
The paper was originally edited as a contribution for  the book
{\bf J. Kertesz and I. Kondor (Editors), Econophysics: an
Emerging Science},
 Kluwer Academic Publishers, Dordrecht (NL) that 
 should contain selected papers presented at that Workshop 
 and should have appeared in 1998 or 1999.
 Unfortunately the book was not published. The present e-print is a revised
 version (with up-date annotations and references) of that unpublished contribution,
 but essentially represents our knowledge of that early time.}

\vvs

 {Francesco MAINARDI}$^{(1)}$,
 {Paolo PARADISI}$^{(2)}$ and
{Rudolf GORENFLO}$^{(3)}$

\vs
$\null^{(1)}$ Department of Physics, University of Bologna, and INFN,\\
Via Irnerio 46, I-40126 Bologna, Italy. \\
{\tt francesco.mainardi@unibo.it} \ {\tt francesco.mainardi@bo.infn.it}
\\ [0.25 truecm]
$\null^{(2)}$
ISAC: Istituto per le Scienze dell'Atmosfera e del Clima
    del CNR, \\ 
 Strada Provinciale Lecce-Monteroni Km 1.200, I-73100 Lecce, Italy.\\
{\tt p.paradisi@isac.cnr.ir}
\\ [0.25 truecm]
$\null^{(3)}$ Department of Mathematics and Computer Science,\\
	Freie Universit\"at  Berlin,
Arnimallee  3, D-14195 Berlin, Germany. \\
{\tt gorenflo@mi.fu-berlin.de}

\end{center}

\input{mainardi_BUDAPEST97_C.tex}
\newpage
\cen{\bf Abstract} 

\vskip 0.1truecm
\noindent
Fractional calculus allows one to generalize
the linear, one-dimensional, diffusion equation  by
replacing either the first time derivative  or the second
space derivative by   a derivative of	fractional order.
The fundamental solutions of these generalized diffusion equations are
shown to provide   probability density functions, 
evolving on time or variable in  space, which are related to the
peculiar class of stable distributions.
This property is a noteworthy
generalization of what happens for 
the standard diffusion equation and  can be relevant in
treating financial and economical problems where the
stable probability distributions are known to play a key role.

\input{mainardi_BUDAPEST97_1.tex}
\input{mainardi_BUDAPEST97_2.tex}
\input{mainardi_BUDAPEST97_3.tex}
\input{mainardi_BUDAPEST97_4.tex}
\input{mainardi_BUDAPEST97_5.tex}
\input{mainardi_BUDAPEST97_6.tex}
\input{mainardi_BUDAPEST97_7.tex}

\input{mainardi_BUDAPEST97_A.tex}
\input{mainardi_BUDAPEST97_B.tex}
\newpage
\input{mainardi_BUDAPEST97_R.tex}
\end{document}

%% file: mainardi_BUDAPEST97_C.tex
 \def\indice{\leaders\hbox to 1 em {\hss.\hss}\hfill}
 \def\hb#1{\hbox to 0.75 truecm{p. \hss#1}}
\centerline{\bf Contents}
\vskip 0.4 truecm
\pni {\ \ \ \ {Abstract}\indice  \hb{2}}
\pni {{1. Introduction}\indice \hb{2}}
\pni {{2. The Standard Diffusion Equation}\indice \hb{4}}
\pni {{3. The Time-Fractional Diffusion Equation}\indice \hb{8}}
\pni {{4. The Cauchy Problem for the Time-Fractional
	  Diffusion Equation}\indice \hb{10}}
\pni {{5. The Signalling Problem for the Time-Fractional
	  Diffusion Equation}\indice \hb{13}}
\pni {{6. The Cauchy Problem for the Symmetric Space-Fractional}}
\pni{\ \ \ \ {Diffusion Equation}\indice \hb{15}}
\pni {{7. Conclusions}\indice	\hb{21}}
\pni {{A. The Riemann-Liouville Fractional Calculus}\indice \hb{22}}
\pni {{B. The Stable Probability Distributions}   \indice   \hb{31}}
\pni {{{\phantom{C.}\ }References} \indice  \hb{41}}

%% file: mainardi_BUDAPEST97_1.tex

\section{Introduction}

Non-Gaussian probability distributions are becoming more common
as data models, especially in economics where large fluctuations
are expected. In fact,	probability distributions with heavy tails  are
often  met in economics and finance, which suggests to enlarge the arsenal
of possible stochastic models by non-Gaussian processes.
This conviction started  in the early sixties  after the appearance of a
series of papers by Mandelbrot and his associates,
who point out the importance of non-Gaussian probability distributions,
formerly introduced by Pareto and L\'evy, and  related scaling properties,
to analyse  economical and financial variables,
as reported in the recent book by Mandelbrot (1997).
Some examples of such variables are common stock prices changes,
changes in other speculative prices, and interest rate changes.
In this respect many  works by different authors have recently appeared,
see \eg the recent books by Bouchaud \& Potter (1997),
Mantegna \& Stanley (1998) and the references therein quoted.
\vsp
It is well known that the fundamental solution (or {\it Green} function)
of the {\it Cauchy} problem for the  standard linear diffusion equation
provides at any time the   probability density function ($pdf$)
in space  of the {\it Gauss} (or normal) law.
This law exhibits all moments finite thanks to its exponential decay at
infinity. In particular, the space variance of the {\it Green} function
is proportional to the first power of time, a noteworthy property that can
be understood by  means of an unbiased random walk model for the Brownian
motion, see \eg Feller	(1957).
Less known is the property  for which the fundamental solution
of the {\it Signalling} problem for the same diffusion equation,
provides at any position a unilateral $pdf$
in time, known as {\it L\'evy} law, using the terminology of Feller
(1966-1973). Because of its algebraic decay at infinity as $t^{-3/2}\,,$
this law has all moments of integer order  divergent, and
consequently its expectation value and variance are infinite.
\vsp
Both the {\it Gauss} and {\it L\'evy} laws
belong to the general class of	{\it stable}
probability distributions, which are characterized by an
index $\alpha $ ($0<\alpha \le 2$), called index of stability or
characteristic exponent. In particular,
the index of the {\it Gauss} law is $2\,, $ whereas that of the
{\it L\'evy} law is $1/2\,.$
\vsp
In this paper  we  consider two different  generalizations of
the diffusion equation by  means of fractional calculus,
which allows us to replace either the first time derivative  or the second
space derivative by a suitable fractional derivative.
Correspondingly, the   generalized equation will be referred to as
the {\it time-fractional} diffusion equation
or the	 {\it symmetric, space-fractional} diffusion equation.
Here we show how the fundamental solutions of this equation for the
{\it Cauchy} and {\it Signalling}
problems  provide probability density functions related
to certain stable  distributions, so providing a
natural generalization	of what occurs for the standard diffusion equation.
\vsp
The plan of the paper is as follows.
First of all, for the sake of convenience and completeness, we provide
the essential notions  of {\it Riemann-Liouville Fractional Calculus} and
{\it L\'evy Stable Probability Distributions} in Appendix A and B,
respectively.
\vsp
In Section 2, we recall the basic results for the standard diffusion equation
concerning
the fundamental solutions
of the {\it Cauchy} and {\it Signalling} problems.
In particular we provide the
derivation  of these solutions by the Fourier and Laplace transforms
and the interpretation in terms of {\it Gauss} and {\it L\'evy} stable
$pdf\,, $ respectively.
\vsp
In Section 3, we  consider the {\it time-fractional} diffusion equation
and we formulate for it  the basic {\it Cauchy} and {\it Signalling}
problems to be treated in the subsequent two sections.
Here we adopt the Riemann-Liouville approach to
Fractional Calculus, and the related definition  for the  Caputo
time-fractional derivative of a causal function of time.
\vsp
In Section 4, we solve the {\it Cauchy} problem for the {\it time-fractional}
diffusion equation by using the technique of Fourier transform and
we derive the corresponding fundamental solution in
terms of a special function of Wright type in the similarity
variable.
In this case
the solution  can be interpreted as a noteworthy symmetric $pdf$
in  space with all moments finite, evolving in time.
In particular, its space variance turns out to be proportional to
a power of time equal to the order of the time-fractional derivative.
\vsp
In Section 5, we derive the fundamental solution for	the {\it Signalling}
problem of the {\it time-fractional}
diffusion equation by using the technique of Laplace transform.
In this case  the  solution,  still
expressed in terms  of a special function of Wright type,
can be interpreted as a unilateral stable $pdf$
in  time, depending on position,
with index of stability given by
half of the order of the time-fractional derivative.
\vsp
In Section 6, we  consider the {\it symmetric, space-fractional} diffusion
equation. Here we adopt the Riesz approach to
Fractional Calculus, and  the related definition
for the  symmetric space-fractional derivative of a function
of a single space  variable.
Here we treat the {\it Cauchy} problem by
technique of Fourier transform and we derive the series
representation of the corresponding {\it Green} function.
In this case the fundamental solution is interpreted
in terms of a symmetric stable $pdf$ in space,
evolving in time, with index of
stability given by the order of the space-fractional derivative.
To approximate such evolution
we propose  a random walk model, discrete in space and time,
which is based on the Gr\"unwald-Letnikov approximation of
the fractional derivative.
\vsp
Finally, Section 7 is devoted to  conclusions and remarks
on  related work. 

%% file: mainardi_BUDAPEST97_2.tex

\section{The standard diffusion equation}

    For the standard diffusion equation
we mean the linear partial differential equation
$$ {\d \over \dt}\, u(x,t) = \D \, {\d^2\over \dx^2}\, u(x,t)\,,
\qq u=u(x,t)\,,\eqno(2.1)$$
where
$\D$ denotes a positive constant with the dimensions $L^2\,T^{-1}\,,$
$x$ and $t$ are the space-time variables,
and  $u=u(x,t)$ is the field variable, which is assumed
to be a {\it causal} function of time, \ie
vanishing for $t<0\,.$


\vsp
The typical physical phenomenon  related  to such an equation is
the heat conduction in a thin solid rod extended along $x\,, $
so the field variable $u$ is the temperature.
\vsp
In order to  guarantee the existence and the uniqueness of the
solution, we must equip (1.1) with suitable data on the
boundary of the space-time domain.
The  basic boundary-value problems
for  diffusion
are the so-called {\it Cauchy} and {\it Signalling} problems.
In the	{\it Cauchy} problem, which concerns  the space-time domain
$-\infty <x< + \infty\,, $ $\, t \ge 0\,, $
the data are assigned at $t=0^+$ on
the  whole space axis (initial data).
In the {\it Signalling} problem, which concerns  the space-time domain
$x\ge 0\,, $ $\, t \ge 0\,, $
the data are assigned both at $t=0^+$ on the semi-infinite
space axis $ x >0 $ (initial data) and at $x=0^+$ on the semi-infinite
time axis  $ t>0$ (boundary data); here, as mostly usual,
the initial data are assumed to be vanishing.
\vsp
Denoting by $g(x)$ and $h(t)$ two given, sufficiently well-behaved
functions,    the basic problems are thus formulated as following:
\vsh\pni
{\it a) Cauchy} problem
$$   u(x,0^+)=g(x) \,, \q -\infty <x < +\infty\,; \q
     u(\mp \infty,t) = 0\,,\q \, t>0\,;  \eqno(2.2a)
$$
\pni
{\it b) Signalling} problem
$$  u(x, 0^+) =0 \,, \q  x>0\,;\q
    u(0^+,t ) =h(t) \,, \q u(+\infty,t) =0 \,, \q   t >0 \,. \eqno(2.2b)
$$
Hereafter, for both the problems,
 we  derive  the classical results which will
be properly generalized for the fractional diffusion equation
in the subsequent sections.
 \vsp
Let us begin with the {\it Cauchy} problem.
It is well known that this initial value  problem
can be easily solved making use of the Fourier transform and its
fundamental solution can be interpreted as a Gaussian $pdf$ in $x$.
Adopting the notation
$ g(x) \div  \hat g(\kappa) $  with $\kappa \in \RR$ and
 $$  \hat g(\kappa) = \F\l[g(x)\r] =
\int_{-\infty}^{+\infty}
   \!\! \e^{\,\ds +i\kappa x}\, g(x)\, dx\,,   $$
$$ g(x) = {\F}^{-1}\l[\hat g(\kappa )\r] = \rec{2\pi}\,
 \int_{-\infty}^{+\infty}
   \!\! \e^{\,\ds -i\kappa x}\, \hat g(\kappa )\, d\kappa \,,
$$
the transformed solution 
satisfies the ordinary differential equation of the first order
$$  \l({d \over dt} + \kappa^2\,\D \r)\, \hat u(\kappa ,t) =0\,,
  \q \hat u(\kappa ,0^+) =\hat g(\kappa )\,, \eqno(2.3)    $$
and consequently it turns out to be
$$ \hat u(\kappa ,t) = \hat g(\kappa) \, \e^{\,\ds -\kappa^2\,\D\,t}\,.
\eqno (2.4)$$
Then, introducing
$$ \Gc^d(x ,t) \div \widehat{\Gc^d}(\kappa ,t)	=
\e^{\,\ds -\kappa ^2\,\D\,t}\,, \eqno(2.5)$$
where the upper index $d$ refers to (standard) diffusion,
the required solution, obtained by inversion of (2.4), can be expressed
in terms of the  space convolution
$ u(x,t)
= \int_{-\infty}^{+\infty} \Gc^d(\xi ,t) \, g(x-\xi  )
\, d\xi     \,,$ where
$$ \Gc^d (x,t)
 = {1\over 2\sqrt{\pi \,\D}}\,t^{-1/2}\, \e^{-\ds\arg}\,.
\eqno(2.6)$$
Here $\Gc^d(x,t)\, $ represents the  fundamental solution
(or {\it Green} function) of the {\it Cauchy} problem, since it
corresponds to $g(x) =\delta (x)\,. $
It turns out to be a function in $x\,, $ even and normalized,
\ie  $\Gc^d(x,t)=  \Gc^d(|x|,t)$ and
$ \int_{-\infty}^{+\infty}  \Gc^d(x,t)\, dx =1\,.$
We also note the identity
$$ |x|\,  \Gc^d(|x|,t)	=  {\zeta \over 2}\, M^d(\zeta) 
 \,, \eqno(2.7)$$
where $\zeta =|x|/(\sqrt{\D}\, t^{1/2})$ is the well-known {\it similarity}
variable and
$$ M^d(\zeta)= \rec{\sqrt{\pi}} \,\e^{-\ds \zeta^2/4}\,.\eqno(2.8)$$
We note that $M^d(\zeta)$ satisfies the normalization condition
$ \int_0^{\infty}  M^d(\zeta)\, d\zeta =1\,.$
\vsp
The interpretation of the {\it Green} function (2.6) in probability theory
is straightforward since we easily recognize
$$  \Gc^d(x,t) =
   p_G(x; \sigma) :=
 {1\over \sqrt{2\pi}\,\sigma }\, \e ^{\,\ds -x^2/(2 \sigma ^2)}\,, \q
  \sigma^2 =2\, \D\, t\,,
 \eqno(2.9) $$
where $p_G(x; \sigma)$ denotes the well-known  {\it Gauss} or {\it normal}
$pdf$ spread out over all real $x$ (the space variable),
whose moment of the second order, the {\it variance}, is $\sigma ^2\,. $
The associated cumulative distribution function ($cdf$) is known to be
$$ {\cal{P}}_G(x;\sigma ) :=
 \int_{-\infty}^x \!\!\! p_G(x';\sigma )\, dx'
  ={\rec 2} \l[ 1 + \erf \, \l( {x\over \sqrt{2}\, \sigma }\r)\r]
 \,,  \eqno(2.10)$$
where $\,\erf\, (z):= (2/\sqrt{\pi})\,\int_0^z {\exp}\,(-u^2)\, du\,$
denotes the error function.
Furthermore,
the  moments of   even order of the {\it Gauss} $pdf$
 turn out to be
$ \int_{-\infty}^{+\infty} x^{2n}\,
  p_G (x;\sigma ) \,dx =  (2n-1)!! \, \sigma ^{2n}  \,, $ so
$$ \int_{-\infty}^{+\infty} \!\!\!\!x^{2n}\,
  \Gc^d (x,t) \,dx
  = (2n-1)!! \, (2\D\,t)^n    \,,\q  n=1,2,\dots\,.\eqno(2.11)$$
\vsp
Let us now consider the {\it Signalling} problem.
This initial-boundary value problem
can be easily solved by making use of the Laplace transform.
Adopting  the notation $h(t) \div \tilde h(s)$ with $s\in \CC$ and
$$ \tilde h(s)	= \L\l[h(t)\r] =
 \int_{0}^{\infty}
   \!\! \e^{\,\ds -s t}\, h(t)\, dt\,, $$
$$ h(t)  = {\L}^{-1}\l[\tilde h(t)\r] =
 \rec{2\pi \,i}\, \int_{Br}
   \!\! \e^{\,\ds s t}\, \tilde h(s)\, ds\,,  $$
where $Br$ denotes the Bromwich path,
the transformed solution of the diffusion equation satisfies
the ordinary differential equation of the second order
$$ \l( {d^2\over dx^2} - {s\over \D}\r) \, \tilde u(x,s)=0\,,
\q \tilde u(0^+, s) = \tilde h(s)\,, \q
 \tilde u(+\infty, s) =  0\,. \eqno(2.12)$$
and consequently it turns out to be
 $$ \tilde u(x,s) = \tilde h(s) \, \e^{\,\ds -(x/\sqrt{\D})\,s^{1/2}}
\,. \eqno (2.13)$$
Then introducing
$$ \Gs^d(x ,t) \div \widetilde \Gs^d(x,s) =
\e^{\,\ds -(x/\sqrt{\D})\,s^{1/2}}\,, \eqno(2.14)$$
the required solution, obtained by inversion of (2.13),  can be expressed
in terms of the time convolution,
 $
u(x,t)= \int_{0}^{t} \Gs^d(x ,\tau) \, h(t-\tau)\, d\tau \,,$
where
$$ \Gs^d (x,t) = {x\over 2\sqrt{\pi \,\D}}\,t^{-3/2}\, \e^{-\ds\arg}\,.
\eqno(2.15)
$$
Here $\Gs^d(x,t)$ represents the fundamental solution
(or {\it Green} function) of the {\it Signalling} problem,
 since it corresponds
to $h(t) = \delta (t)\,. $ We note that
$$
 \Gs^d (x,t) =	 p_{LS}(t; \mu ):=
   {\sqrt{\mu }\over \sqrt{2\pi}\, t^{3/2}}\, \e ^{\,\ds -\mu /(2t)}\,,
 \q  t\ge 0\,, \q      \mu =  {x^2\over 2\,\D}\,,
\eqno(2.16)
$$
where $p_{LS}(t;\mu )$
denotes the  {\it one-sided} {\it L\'evy-Smirnov} $pdf$ spread out over
all non negative $t$ (the time variable).
The associated $cdf$ is,
see \eg Feller (1966-1971) and Pr\"uss (1993),
$$ {{\cal P}}_L(t;\mu ) := \int_0^t \!\! p_L(t' ;\mu  )\, dt'  =
    {\erfc}\, \l( \sqrt{{\mu \over 2t}}\r)  =
 {\erfc} \,\l( {x\over 2\,\sqrt{\D\,t}}\r)
      \,,   \eqno(2.17)$$
where $\erfc\, (z) := 1 - \erf\, (z)$
denotes the complenatary error function.
\vsp
The {\it L\'evy-Smirnov} $pdf$ has all moments  of integer order infinite,
since it decays at infinity as $t^{-3/2}\,. $ However,
we note that
the absolute moments of real order $\nu $ are finite only if
$ 0\le \nu   <1/2\,. $	In particular, for this $pdf$  the
mean  is infinite, for which we can take  the {\it median} as
expectation value.
From ${{\cal P}}_{Ls}(t_{med};\mu)=1/2\,, $ it turns out that
$t_{med} \approx  2 \mu \,, $ since the complementary
error function gets the value 1/2 as its argument is
approximatively 1/2.


\vsp
We  note that in the common domain $x>0\,, \, t>0$
the Green functions of the two basic problems
satisfy the  identity
$$ x\,	\Gc^d(x,t) = t\, \Gs^d(x,t)\,,\eqno(2.18)$$
that we refer to as the {\it reciprocity relation} between the
two fundamental solutions of the diffusion equation.
Furthermore, in view of (2.7) and (2.18)  we recognize
the role of the function   of the similarity variable,
 $M^d(\zeta)\,, $
in providing the two fundamental solutions;
we shall refer to it as to
the {\it normalized auxiliary function} of the diffusion equation
for both the {\it Cauchy} and {\it Signalling} problems.


%% file: mainardi_BUDAPEST97_3.tex

\section{The time-fractional diffusion equation}

By the {\it time-fractional diffusion} equation we mean the  linear
evolution equation obtained from the
classical diffusion  equation by replacing the first-order time
derivative by  a fractional derivative (in the {\it Caputo} sense)
of order  $\alpha   $ with $0 <\alpha	\le 2$.
In our notation it reads
$${\d^{\alpha }  u\over \dt^{\alpha  }} =
 \D\,{\d^2 u \over \dx^2} \,,\qq u=u(x,t)\,, \q
0<\alpha \le 2 \,, \eqno(3.1)$$
where
$\D$ denotes a positive constant with the dimensions $L^2\,T^{-\alpha }\,.$
From  Appendix A
we  recall the definition of the {\it Caputo}  fractional derivative
of order  $\alpha >0 $
for a (sufficiently well-behaved) {\it causal} function $f(t)\,, $
see (A.9),
$$ D_*^\alpha \, f(t) :=
     {1\over \Gamma(m-\alpha)}\,
 \int_0^t (t-\tau )^{m-\alpha }\, f^{(m)}(\tau )\, d\tau \,,
  \eqno(3.2)
$$
where  $m =1,2, \dots \,, $
and   $0 \le m-1<\alpha \le m\,.$
According to (3.2)
we thus need to distinguish the cases
 $ 0<\alpha \le 1\,$  and $1<\alpha \le 2\,. $
In the the latter case (3.1) may be seen as
a sort of interpolation between the standard diffusion equation
and the standard wave equation.
Introducing
$$ \Phi_{\lambda}(t) := { t_+^{\lambda	-1}\over  \Gamma(\lambda )}\,,\q
   \lambda   >0\,,\eqno(3.3) $$
where the suffix $+$ is just denoting that the function is
vanishing for $t<0\,, $ we easily recognize
that the equation  (3.1)
assumes the explicit forms :
\pni
if $ 0<\alpha \le 1\,, $
$$ \Phi_{1-\alpha} (t) \,*\, {\d u\over \dt} =
    \rec{\Gamma(1-\alpha)}\,
 \int_0^t (t-\tau )^{-\alpha}\, \l( {\d u\over \d\tau}\r) \, d\tau
   =   \D\,{\d^2 u \over \dx^2}  \,; \eqno(3.4)$$
if $1<\alpha \le 2\,, $
$$ \Phi_{2-\alpha} (t) \,*\, {\d^2 u\over \dt^2} =
    \rec{\Gamma(2-\alpha)}\,
 \int_0^t (t-\tau )^{1-\alpha} \, \l({\d^2 u\over\d\tau^2}\r)\, d\tau
   =   \D\,{\d^2 u \over \dx^2}  \,. \eqno(3.5)$$


\vsp
Extending
the classical analysis for the standard diffusion equation (2.1)
to the above integro-differential equations (3.4-5),
the {\it Cauchy} and {\it Signalling} problems are thus formulated as
in equations (2.2), \ie
\vsh\pni
{\it a) Cauchy} problem
$$   u(x,0^+)=g(x) \,, \q -\infty <x < +\infty\,; \q
     u(\mp \infty,t) = 0\,,\q \, t>0\,;  \eqno(3.6a)
$$
\pni
{\it b) Signalling} problem
$$  u(x, 0^+) =0 \,, \q  x>0\,;\q
    u(0^+,t ) =h(t) \,, \q u(+\infty,t) =0 \,, \q   t >0 \,. \eqno(3.6b)
$$
However,
if $1<\alpha   \le 2\,, $ the presence in (3.5)
of the second order
time derivative of the field variable  requires to specify
the initial value of the first order time derivative
$u_t(x,0^+)\,,$ since in this case   two linearly independent
solutions are to be determined.
To ensure the continuous dependence  of our solution
on the parameter $\alpha   $
also in the transition from $\alpha   =1^-$ to	$\alpha   =1^+\,,$
we agree  to  assume $u_t(x,0^+) = 0\,. $
\vsp
We  recognize that our fractional diffusion equation (3.1),
when subject to the conditions (3.6), is equivalent
to the integro-differential equation
$$  u(x,t)
= g(x) + {\D\over \Gamma(\alpha)}\,
  \int_0^t  (t-\tau)^{\alpha-1}\, \l({\d^2 u\over \dx^2}\r) \, d\tau\,,
    \eqno(3.7)$$
where $ 0<\alpha \le 2 \,.$
Such integro-differential equation has been investigated by several
authors, including Schneider \& Wyss (1989), Fujita (1990), Pr\"uss (1993)
and  Engler (1997).
\vsp
In view of our subsequent analysis we find it convenient to put
$$  \nu  ={\alpha \over 2}\,,
   \q 0<\nu  <1\,.  \eqno(3.8)$$
In fact  the analysis of
the {\it time-fractional} diffusion equation turns out to be
easier if we adopt as a key parameter the half of the order
of the {\it time-fractional} derivative. In future we shall
provide  the symbol $\alpha $ with other relevant meanings,
as  the index of stability of a stable probability distribution
or the order of the  space derivative in the
{\it space-fractional} diffusion equation.
\vsp
Henceforth, we agree to insert the parameter $\nu $ in the
field variable, \ie  $u=u(x,t;\nu )\,. $
By  denoting  the Green functions
of the {\it Cauchy} and {\it Signalling}
problems
by $\Gc (x,t;\nu )$ and $\Gs(x,t;\nu )\,, $
respectively,
the solutions of the two basic problems
are obtained by a space or time convolution,
$ u(x,t;\nu)
= \int_{-\infty}^{+\infty} \Gc(\xi ,t;\nu) \, g(x-\xi)
\, d\xi \,, $
$ u(x,t;\nu) = \int_{0}^{t} \Gs(x,\tau;\nu)\, h(t-\tau) \,d\tau\,,$
respectively.
It should be noted that  $\Gc(x ,t ;\nu) =  \Gc(|x|,t ;\nu)\,,$
since the Green function turns out to be
an even function of $x\,.$
\vsp
In the following two sections we shall compute
the two fundamental solutions
with the same techniques (based on Fourier and Laplace transforms)
used for the standard
diffusion equation and we shall provide their interpretation in terms of
probability distributions.
Most of the presented results are based on the papers by Mainardi
(1994), (1995), (1996), (1997) and by Mainardi \& Tomirotti
(1995), (1997).

%% file: mainardi_BUDAPEST97_4.tex

\section{The Cauchy problem for the time-fractional diffusion
    equation}

For the fractional diffusion  equation (3.1) subject to (3.6a)
the application of the Fourier transform leads
to the ordinary differential equation of order	$\alpha= 2\nu \,,$
$$  \l({d^{2\nu }   \over dt^{2\nu }  } + \kappa^2 \,\D\r)\, \hat
u(\kappa, t;\nu  )    =0\,,
 \q \hat u(\kappa ,0^+;\nu  ) =\hat g(\kappa )\,, \eqno(4.1)	$$
Using the results of Appendix A, see (A.22-30),
 the transformed solution is 
$$ \hat u(\kappa ,t;\nu  ) = \hat g(\kappa)
  \, E_{2\nu }	\l( -\kappa^2\,\D\,t^{2\nu } \r)\,, \eqno (4.2)$$
where $E_{2\nu }  (\cdot)$ denotes the {\it Mittag-Leffler} function
of order $ 2\nu   \,, $
and
consequently for the  {\it Green} function we have
$$ \Gc(x,t;\nu )  = \Gc(|x|,t;\nu )
\div \hat{\Gc}(k,t;\nu	) =
     E_{2\nu  } \l( -\kappa^2\D\,t^{2\nu }  \r)\,.\eqno(4.3)$$
\vsp
Since the {\it Green} function is a real and even function of $x$,
its (exponential) Fourier transform can be  expressed in terms of the
{\it cosine} Fourier transform and thus is related to its spatial Laplace
transform  as follows
$$
\begin{array}{ll}
$$ \hat{\Gc}(k,t;\nu) \;
 =&{\ds 2\,\int_0^\infty\!\! \Gc(x,t;\nu)\,\cos\, \kappa x\,dx} \,= \\ \ \\
 \null &\l.\tilde{\Gc}(s,t;\nu)\r\vert_{s=+ik}
   + \l.\tilde{\Gc}(s,t;\nu)\r\vert_{s=-ik}\,.
\end{array}
 \eqno(4.4) $$
Indeed, a split  occurs also in (4.3) according to the {\it duplication
formula} for the {\it Mittag-Leffler} function, see (A.26),
  $$  \begin{array}{ll}
   \hat{\Gc}(k,t;\nu) \;
  =&E_{2\nu  }(-\kappa^2\,\D\,t^{2\nu }) \,= \\ 
\null &[E_{\nu}(+i\kappa\,\sqrt{\D}\,t^{\nu})
   + E_{\nu} (-i\kappa\,\sqrt{\D}\,t^{\nu})]/2 \,.
  \end{array}
\eqno(4.5)$$
When $\nu  \ne 1/2$ the inversion of the Fourier transform in (4.5)
 cannot be obtained by using a standard table of Fourier transform pairs;
 however, for any $\nu \in (0,1)$ such inversion can be achieved
by  appealing to  the Laplace transform pair (A.37) with $r= |x|\,, $  and
$s = \pm i\kappa \,.$
In fact, taking into account the scaling property of the Laplace transform,
we  obtain from (4.5) and (A.37)
$$ \Gc(|x|,t;\nu ) = \rec{2\, \sqrt{\D}\, t^\nu  }\,
    M\l({|x|\over \sqrt{\D}t^\nu };\nu	\r) \,, \eqno(4.6)$$
where $M(\zeta ;\nu )$ is the special function
of Wright type, defined by (A.31-33),
and
$$ \zeta  = {|x|\over \sqrt{\D} t^\nu }\,, \eqno(4.7)$$
the  {\it similarity} variable.
We note the identity
$$  |x|\,  \Gc(|x|,t;\nu ) = {\zeta \over 2}\, M(\zeta;\nu ) \,,
\eqno(4.8)$$
which generalizes to the time-fractional  diffusion equation
the identity (2.7) of the standard diffusion equation.
Since $\int_0^\infty M(\zeta;\nu )\, d\zeta =1\,, $ see (A.40),
the function $M(\zeta ;\nu )\, $
is  the {\it normalized auxiliary} function of the fractional diffusion
equation.
\vsp
We note that for the time-fractional diffusion equation the  fundamental
solution of the {\it Cauchy} problem is still  a bilateral symmetric $pdf$
in $x$ (with two 
branches, for $x>0$ and $x<0\,, $
obtained one from the other by reflection), but is
no longer of Gaussian type if $\nu  \ne 1/2\,. $
In fact, for large $|x|$ each branch exhibits an exponential decay in the
"stretched" variable $|x|^{1/(1-\nu )}$
as can be
derived from the asymptotic representation (A.36) of the {\it auxiliary}
function $M(\cdot;\nu )\,. $
In fact,  by using (4.7-8) and (A.36),
we obtain
 $$  \Gc(x,t;\nu  ) \sim a_*(t)\, |x|^{(\nu -1/2)/(1-\nu )}
   \, {\exp}\, \l[-b_*(t) |x|^{1/(1-\nu )}\r]\,,
\eqno(4.9)$$
as $ |x|\to \infty\,,$
where  $a_*(t)$ and $b_*(t)$ are certain positive functions of time.
\vsp
Furthermore, the exponential decay in $x$ provided by (4.9) ensures that
all the absolute moments  of positive order of $\Gc(x,t;\nu )$ are finite.
In particular,	  using  (4.8) and (A.39)
it turns out
that the moments (of even order) are
$$ \int_{-\infty}^{+\infty} \!\!\! x^{2n}\, \Gc(x,t;\nu  )  \,dx =
   {\Gamma(2n+1)\over \Gamma(2\nu  n+1)}\, (\D t^{2\nu	})^n\,,
\q n=0\,, \,1\,, \,2\,, \, \dots
\eqno(4.10)$$
The formula (4.10) provides  a generalization of the corresponding
formula (2.11) valid for the standard diffusion equation, $\nu	=1/2\,.$
Furthermore, we  recognize that the variance associated to the $pdf$ is
now proportional to $\D t^{2\nu  }\,, $ which for $\nu \ne 1/2$ implies a
phenomenon of {\it anomalous} diffusion. According to a usual terminology
in statistical mechanics, the anomalous diffusion is said to be    slow if
$0<\nu	<1/2$ and fast	if $1/2<\nu  <1\,. $

\vsp
In Figure 1, as an example,
we compare versus $|x|\,, $
at fixed $t\,, $  the fundamental solutions of the {\it Cauchy} problem
with different $\nu $ ($\nu  =1/4\,, \, 1/2\,, 3/4\,$).
We consider the range $0\le |x|\le 4$ and assume $\D=t =1\,. $
\vsp
 \begin{figure}[h]
\centering
 \includegraphics[width=.60\textwidth]{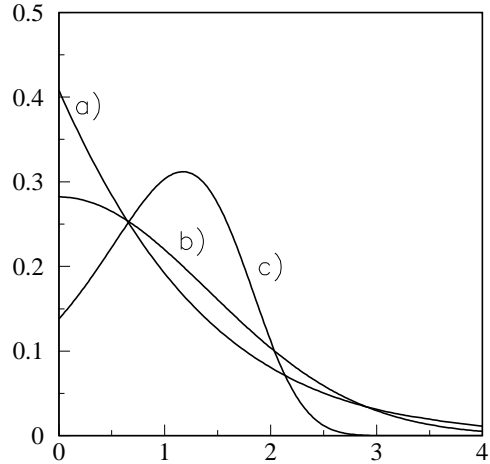}
 \vskip -0.5truecm
\caption{%
 The Cauchy problem for the time-fractional diffusion equation. 
 The fundamental solutions versus $|x|$  with
 a) $\nu =1/4\,, \;$	 b) $\nu =1/2\,, \;$  c) $\nu
=3/4\,. $} 
\label{buda1.fig}  
\end{figure}
\vsp
We note the different  behaviour of the $pdf$
in the cases of slow diffusion ($\nu  = 1/4\,$)
and fast  diffusion ($\nu  =3/4\,$)
with respect to the Gaussian
behaviour of the standard diffusion ($\nu  =1/2$).
In the limiting cases $\nu  = 0$ and $\nu  =1$	we
have
$$
\Gc(x,t;0) = {\ds{{\e}^{\,\ds -|x|}\over 2}}\, , \q
\Gc(x,t;1) =
{\ds{\delta (x-\sqrt{\D}\,t)+\delta (x+\sqrt{\D}\,t)\over 2}}\,.
  \eqno(4.11)
$$
\vsp
We also recognize from the appendix B that for $1/2 \le \nu  <1$
any branch of the fundamental solution is proportional to the
corresponding {\it positive}
branch of an extremal stable $pdf$ with index of stability
$\alpha =1/\nu	\,, $ which exhibits an exponential decay at infinity.
In fact, applying (B.29)
with $\alpha =1/\nu  \, $ and	$y= \zeta= |x|/(\sqrt{\D}t^\nu)\,,$
from (4.7-8) we obtain
$$
\begin{array}{lll}
  \Gc(|x|,t;\nu)\,
  =& {\ds{1\over 2\, \sqrt{\D}\,t^\nu}}
  \cdot {\ds{\rec{\nu}}}\,
      q_{1/\nu }\, \l[ {|x|/(\sqrt{\D}\,t^\nu)\,;\, -\,(2-1/\nu)}\r] \,=
  \\ \ \\
   \null &{\ds\rec{2\nu}} \cdot p_{1/\nu}\,(|x|;\,+1,1,0) \,,
    \qq 1<1/\nu \le 2\,.
    \end{array}
\eqno(4.12) $$
We also note that the  stable distribution in (4.12)
 satisfies the condition
   $$ \int_{0}^{+\infty} \!\!p_{1/\nu} \,(x;\,+1,1,0) \,dx =
	 \nu \,, \qq  1 < 1/\nu  \le 2\,. \eqno (4.13)$$



%% file: mainardi_BUDAPEST97_5.tex
\section{The Signalling problem for
 the time-fractional diffusion equation}

For the fractional diffusion  equation (3.1) subject to (3.6b)
the application of the Laplace transform leads
to the ordinary differential equation of order	$ 2\,,$
$$  \l({d^{2}	\over dx^{2}} - {s^{2\nu }\over \D} \r)\,
 \tilde u(x,s;\nu )\,,
\q \tilde u(0^+, s;\nu ) = \tilde h(s)\,, \q
 \tilde u(+\infty, s;\nu ) =  0\,. \eqno(5.1)$$
Thus the transformed solution reads
 $$ \tilde u(x,s;\nu ) = \tilde h(s) \, \e^{\,\ds
-(x/\sqrt{\D})\,s^{\nu }} \,, \eqno (5.2)$$
so for the  {\it Green} function we have
$$ \Gs(x,t;\nu ) \div \tilde{\Gs}(x,s;\nu  ) =
  \e^{\,\ds -(x/\sqrt{\D})\,s^{\nu }}	\,.\eqno(5.3)$$
\vsp
When $\nu  \ne 1/2$ the inversion of this Laplace transform  cannot be
obtained by looking in a standard table of Laplace transform pairs.
Also here we appeal to a Laplace transform pair related
 to  the Wright-type  function $M(\zeta ;\nu )$.
In fact, using (A.40) with  $r =t\,, $
and taking into account the scaling property of the Laplace transform,
we  obtain
$$ \Gs(x,t;\nu	) =
  \nu \,   {x\over \sqrt{\D} \,t^{1+\nu }}\,
  M\l({x\over \sqrt{\D}\, t^\nu }  ;\nu \r)\,.	\eqno(5.4)$$
Introducing the  {\it similarity} variable
$ \zeta  = {x/(\sqrt{\D} t^\nu )}\,,$ we  recognize
the identity
$$ t\, \Gs(x,t;\nu  ) = \nu  \, \zeta \, M(\zeta;\nu )\,,
  \eqno(5.5)$$
which is the counterpart  for the {\it Signalling} problem
of the identity (4.8)  valid for the {\it Cauchy} problem.
\vsp
Comparing (5.5) with (4.8) we obtain the {\it reciprocity relation}
between the two fundamental solutions of the {\it time-fractional}
diffusion equation, in the common domain  $x>0\,, \, t>0\,,$
$$ 2\nu  \, x\,  \Gc(x,t;\nu ) = t\, \Gs(x,t;\nu )\,.\eqno(5.6)$$
\vsp
The  interpretation of	$ \Gs(x,t;\nu  )$ as a one-sided
{\it stable} $pdf$ in time is straightforward: in this respect we need to
apply	(B.28), with index of stability $\alpha =\nu $	 and
 variable $y=\zeta^{-1/\nu }= t \, (\sqrt{\D}/x)^{1/\nu }  \,, $   in
(5.5).	 We obtain
$$  \Gs(x,t;\nu  ) =  \l({\sqrt{\D}\over x}\r)^{1/\nu}\,
     q_\nu  \l [ t \, \l({\sqrt{\D}\over x}\r)^{1/\nu}\!\!;\, -\,\nu \r]
  = p_\nu\,  (t;\, -1,1,0)\,. \eqno(5.7)$$
\vsp
In Figure 2, as an example,
we compare versus $t\,, $
at fixed $x\,, $  the fundamental solutions of the {\it Signalling} problem
with different $\nu $ ($\nu  =1/4\,, \, 1/2\,, 3/4\,$).
We consider the range $0\le t\le 3$ and assume $\D=x =1\,. $
\vsp
\begin{figure}[h]
\centering
\includegraphics[width=.60\textwidth]{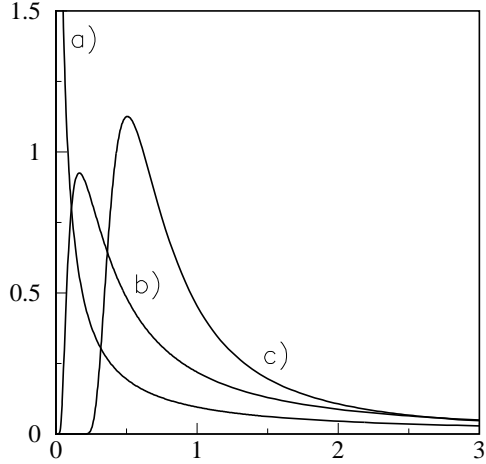}
\vskip -0.5truecm
\caption{The Signalling problem for the time-fractional diffusion
equation. The fundamental solutions versus $t$	with
 a) $\nu =1/4\,, \;$	b) $\nu =1/2\,, \;$  c) $\nu =3/4\,. $}
 \label{buda2.fig}
 \end{figure}
\vsp
We note the different  behaviour of the $pdf$
in the cases of slow diffusion ($\nu  = 1/4\,$)
and fast  diffusion ($\nu  =3/4\,$)
with respect to  the  {\it L\'evy} $pdf$
for the standard diffusion ($\nu  =1/2$).
In the limiting cases $\nu  = 0\,,\,1\,,$	we have
$$\Gs(x,t;0) = \delta (t)\, , \qq
\Gs(x,t;1) = \delta (t- x/\sqrt{\D})\,.
  \eqno(5.8)
$$

%% file: mainardi_BUDAPEST97_6.tex

\section{The Cauchy problem for
 the symmetric space-fractional diffusion equation}

The {\it symmetric space-fractional diffusion} equation
is obtained from the
classical diffusion  equation by replacing the second-order space
derivative by a symmetric space-fractional derivative
(explained below) of order $\alpha$ with
$0<\alpha \le 2\,. $
In our notation we write  this equation as
 $$ {\d \,u \over \dt} = \D
 \, {\d^\alpha\,u \over \d |x|^\alpha}\,, \q u=u(x,t;\alpha)\,,
   \;x \in \RR\,,\; t\in \RR_0^+\,,  \q  0 <\alpha \le 2 \,,
\eqno(6.1)$$
where $\D$ is a positive
coefficient  with the dimensions $L^\alpha \, T^{-1}\,. $
The fundamental solution for the {\it Cauchy} problem,
$\Gc(x,t;\alpha)$
is the solution of (6.1),
subject to the initial condition $u(x, 0^+;\alpha) = \delta(x)\,. $


\vsp
The  {\it symmetric space-fractional} derivative of any order $\alpha >0$
of a sufficiently well-behaved function $\phi(x)\,, \; x\in \RR\,, $
may be defined as   the pseudo-differential
operator characterized in its Fourier representation
by
  $$ {d^\alpha \over d |x|^\alpha}\, \phi(x) \, \div \,
	- |\kappa |^\alpha \,  \hat \phi(\kappa)\,,
    \q x\,,\, k \in \RR \,, \q \alpha >0  \,. \eqno(6.2)$$
 According to a usual terminology, $-|\kappa |^\alpha $ is
 referred to as the symbol of our pseudo-differential  operator,
the {\it symmetric space-fractional} derivative, of order $\alpha \,. $
Here, we have adopted
the notation introduced by Zaslavski, see \eg Saichev \& Zaslavski (1997).
\vsp
In order to  properly introduce this kind of fractional derivative
we need to consider a peculiar approach to  fractional calculus
different from the {\it Riemann-Liouville} one, already treated
in Appendix A. This  approach is indeed based
on the so-called  {\it Riesz} potentials (or integrals),
that  we prefer to consider later.
\vsp
At first, let us see how  things become highly transparent by using  an
heuristic argument, originally due to {Feller} (1952).
 The idea is to start from the
 positive definite differential operator
 $$ A := -{d^2\over dx^2}\,\div \,\kappa^2 = |\kappa |^2\,, \eqno(6.3)$$
 whose symbol is $|\kappa|^2\,, $
and form positive powers of this operator as pseudo-differential
 operators by  their action in the Fourier-image space, \ie
 $$ A^{\alpha /2} := \l(- {d^2\over dx^2}\r)^{\alpha /2}
  \, \div \,  \l(|\kappa |^2\r)^{\alpha /2} = |\kappa |^\alpha
  \q \alpha >0 \,.
 \eqno(6.4)$$
 Thus the operator $ -A^{\alpha /2}$ can be interpreted
 as the  required fractional derivative, \ie
$$  A^{\alpha /2} \equiv -{d^\alpha  \over d|x|^\alpha}  \,, \qq
   \alpha >0\,. \eqno(6.5)$$
 We note that the operator just defined must not be confused with a power
 of the first order differential operator ${d\over dx}$ for which the
symbol is $-i\kappa \,. $
\vsp
After the above considerations it is straightforward to obtain
the Fourier image of the {\it Green} function of the {\it Cauchy}
problem for the {\it space-fractional} diffusion equation.
In fact, applying the Fourier transform to the equation (6.1),
subject to the initial condition $u(x, 0^+;\alpha) = \delta(x)\,, $
and accounting for (6.2), we obtain
$$ \Gc(x,t;\alpha)  = \Gc(|x|,t;\alpha )
\div \hat{\Gc}(k,t;\alpha) = \e^{\, \ds -\D\,t\, |\kappa |^\alpha }\,,
 \q 0<\alpha \le 2\,. \eqno(6.6)$$
We easily recognize that the Fourier transform of the Green function
corresponds to the canonic form of a {\it symmetric} stable distribution
of index of stability $\alpha $ and scaling factor
$\gamma =(\D   t)^{1/\alpha} \,, $ see (B.8).
Therefore we have
$$  \Gc(x,t;\alpha)
   =  p_{\alpha}(x;0,\gamma ,0)\,, \q
  \gamma =(\D	t)^{\, \ds 1/\alpha}  \,. \eqno(6.7) $$
For $\alpha =1$ and $\alpha =2$ we easily obtain the explicit expressions
of the corresponding Green functions since in these cases they
correspond to the {\it Cauchy} and {\it Gauss} distributions,  
   $$  \Gc(x,t;1) = \rec{\pi}\, {\D\,t \over x^2 + (\D\,t)^2}\,,
\eqno(6.8)$$
see  (B.5), and
$$ \Gc (x,t;2))
 = {1\over 2\sqrt{\pi \,\D\,t}}\, \e^{-\ds\arg}\,,
\eqno(6.9)$$
in agreement with (2.6).
\vsp
We easily recognize that
$$   \eta := {|x| \over (\D\,t)^{1/\alpha }} \eqno(6.10)$$
is the {\it similarity} variable for the {\it space-fractional}
diffusion equation, in terms of which we can express the {\it Green}
function for any $\alpha \in (0,2]\,. $  Indeed,
we recognize that
$$  \Gc(x,t;\alpha)
   = \rec{(\D\,t)^{1/\alpha}}\,  q_{\alpha}(\eta;0)  \,, \eqno(6.11) $$
where $q_\alpha (\eta ;0)$ denotes the {\it symmetric} stable
distribution of order $\alpha $ with  Feller-type
characteristic	 function, see (B.14-15).  Now
we can express the Green function
using  the Feller series expansions (B.21-22)
with $\theta =0\,. $
We obtain:
\pni for $0<\alpha <1\,, $
$$
   q_\alpha (\eta ;0) =
- {\ds {1\over \pi\,\eta}}\,  {\ds \sum_{n=1}^{\infty}}
    {\ds {\Gamma (n\alpha +1)\over n!}}\,
  \sin\,\l (n {\ds{\alpha \pi\over 2}}\r)\,
 \l(-\eta^{\ds\,-\alpha}\r)^{\ds n}\,,
      \eqno(6.12a)
$$
\pni for $1<\alpha \le 2\,, $
$$  q_\alpha (\eta ;0) =
  {\ds {1\over \pi\,\alpha}}\,	{\ds\sum_{m=0}^{\infty}}
   (-1)^{\ds m}\, {\ds {\Gamma [(2m+1)/\alpha]\over (2m)!}}
  \, \eta^{\ds	 2m}\,. \eqno(6.12b)
$$
In the limiting case  $\alpha =1$ the above series  reduce to
geometrical series    and therefore are no longer convergent in all of
$\CC\,. $ In particular, they represent the  expansions
of the function $q_1(\eta  ;0) = 1/[\pi (1+ \eta^2)]\,, $
convergent for $ \eta  >1\,$ and $ 0<\eta  <1\,, $ respectively.
\vsp
We also note that for any  $\alpha \in (0,2]\,$
the functions $q_\alpha (\eta ;0)$  exhibit at the origin
the value
 $  q_\alpha (0 ;0 ) = \Gamma(1/\alpha)/(\pi\,\alpha)\,,$
and at the queues, excluding the Gaussian case $\alpha =2\,, $
the {\it algebraic} asymptotic behaviour,
as $ \eta \to \infty\,,$
$$ q_\alpha (\eta ;0) \sim
  \rec{\pi}\, \Gamma(\alpha+1)\, \sin\,\l(\alpha {\pi\over 2}\r)
  \, \eta ^{-(\alpha+1)}   \,,\q
    0<\alpha <2 \,.  \eqno(6.13)$$
\vsp
In Figure 3, as an example,
we compare versus $x\,, $
at fixed $t\,, $  the fundamental solutions of the {\it Cauchy} problem
with different $\alpha $ ($\alpha= 1/2\,, 1\,, \, 3/2\,, 2\,$).
We consider the range  $-6\le x\le +6$ and assume $\D=t =1\,. $
\vsp
\begin{figure}[h]
\centering
\includegraphics[width=.90\textwidth]{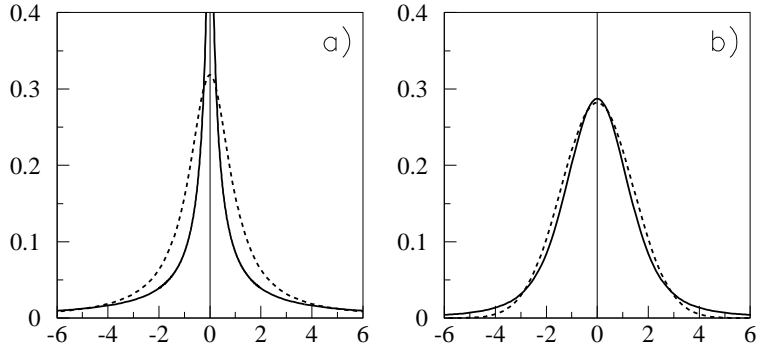}
\vskip-0.5truecm
\caption{The Cauchy problem for the simmetric space-fractional diffusion
 equation. The fundamental solutions versus $x$	:
 plate a) $\alpha=1/2\,$
(continuous line), $\alpha=1\,$ (dashed line);
 plate b) $\alpha=3/4\,$
(continuous line), $\alpha=2\,$ (dashed line).}
 \label{buda3.fig}
\end{figure}
\vsp
Let us now express more properly our operator (6.4) (with
symbol $|\kappa| ^\alpha$)
 as  inverse
of a suitable integral operator $I^\alpha $ whose symbol is
$|\kappa |^{-\alpha }\,. $
This operator can be found in the approach by Marcel Riesz to
Fractional Calculus,
see \eg  {Samko, Kilbas \&  Marichev} (1987-1993) and {Rubin} (1996).
\vsp
We recall that for any
$\alpha >0\,,$ $\, \alpha  \ne 1\,,\,3\,,\, 5 \,,\dots\,$
and for a sufficiently well-behaved function
$\phi(x)\,, \; x\in \RR\,, $
the {Riesz} integral
or {Riesz} potential
$I^\alpha$ and its image
in the Fourier domain read
$$ I^\alpha \, \phi(x) :=
\rec{2\, \Gamma(\alpha )\, \cos(\pi\alpha /2)}\,
\int_{-\infty}^{+\infty}
     \!\!\! |x-\xi |^{\alpha -1}\,\phi(\xi )\, d\xi
  \, \div \, {\hat \phi(\kappa )\over |\kappa |^\alpha} \,.
 \eqno(6.14)$$
On its turn, the {\it Riesz} potential can be written in terms of
two {\it Weyl} integrals  $I^{\alpha }_\pm$ according to
 $$ I^\alpha \, \phi (x) =     \rec{2\, \cos(\pi\alpha /2)}\,
   \l[I^{\alpha }_+ \phi(x) +
   I^{\alpha }_- \phi(x)\r]\,, \eqno(6.15)$$
where
$$ \null      \cases{
   I^{\alpha}_+ \, \phi(x) := {\ds \rec{\Gamma(\alpha)}} \,
 {\ds \int_{-\infty}^{x} \!\!(x-\xi)^{\alpha -1}\,\phi(\xi)\, d\xi}\,,
   & \null \cr\cr
I^{\alpha}_- \,  \phi(x) := {\ds \rec{\Gamma(\alpha)}} \,
{\ds \int_{x}^{+\infty} \!\!(\xi-x)^{\alpha -1}\,\phi(\xi)\,d\xi}\,.
  & \null \cr}
  \eqno(6.16)$$
Then, at least in a formal way, the
 {\it space-fractional} derivative (6.2)
turns out to be defined as the	opposite of the
(left) inverse of the Riesz fractional integral, \ie
$$  {d^\alpha \over d|x|^\alpha }\, \phi (x) :=
 -I^{-\alpha} \, \phi(x) =   -
\rec{2\, \cos(\pi\alpha /2)}\, \l[I^{-\alpha }_+ \phi(x) +
   I^{-\alpha }_- \phi(x)\r]\,. \eqno(6.17)$$
Notice that (6.14) and (6.17) become meaningless when $\alpha $
is an integer odd number.  However, for our range of interest
$0<\alpha \le 2\,, $ the particular case $\alpha =1$ can be
singled out since the corresponding
{\it Green} function is already known, see (6.8).
Thus, excluding the case $\alpha =1\,, $  our {\it space-fractional}
diffusion equation (6.1) can be re-written,
$x\in \RR\,,$	$t \in \RR_0^+\,,$  as
$$ {\d u\over \dt} = - \D\, I^{-\alpha} \, u\,, \q u=u(x,t;\alpha )\,, \q
  0<\alpha \le 2\,, \; \alpha \ne 1\,,
\eqno(6.18)$$
where the operator  $I^{-\alpha}$ is defined by (6.16-17).
\vsp
Here, in order to  evaluate the fundamental solution  of the {\it Cauchy}
problem, interpreted as a probability density,
 we propose a numerical  approach,
original as far as we know,	 based
on  a (symmetric) {\it random walk}
model, discrete in space and time, see also Gorenflo \& Mainardi (1998a),
Gorenflo \& Mainardi (1998b) and Gorenflo, De Fabritiis \& Mainardi (1999).
 We shall
see how things become highly transparent, in that we properly
generalize the classical random-walk argument of the standard
diffusion equation  to our {\it space-fractional}
diffusion equation (6.18).
So doing we are in position to provide  a numerical
simulation of the related (symmetric) stable distributions in a way
analogous  to the standard one for the Gaussian law.
\vsp
The {\it essential idea} is to approximate the left inverse operators
$I_\pm^{-\alpha}$  by the Gr\"unwald-Letnikov scheme, on which the
reader can inform himself in the treatises
on fractional calculus, see \eg
Oldham \& Spanier (1974), Samko, Kilbas \& Marichev (1987-1993),
 Miller \& Ross (1993),
or in the recent review article by Gorenflo (1997).
If $h$ denotes a "small" positive
step-length, these approximating operators read
 $$  _hI^{-\alpha}_\pm \,\phi(x) := \rec{h^\alpha} \,
   \sum_{k=0}^{\infty} (-1)^k\,{\alpha \choose k}\,\phi (x \mp kh)\,.
\eqno(6.19)$$
\def\alphak{{\alpha \choose k}}
\def\alphazero{{\alpha \choose 0}}
\def\alphaone{{\alpha \choose 1}}
\def\alphatwo{{\alpha \choose 2}}
\def\alphakk{{\alpha \choose k+1}}
 Assume, for simplicity,
$\D=1\,, $ and introduce grid points
$\,x_j = j \, h\, $ with $h>0\,,$ $\, j \in \ZZ\,, $
and time instances $\,t_n =n\,\tau\, $ with
$\tau >0\,, \, n\in \NN_0\,. $
Let there be given probabilities $\, p_{j,k}\ge 0\, $
of jumping from point $x_j$ at instant $t_n$ to point $x_k$
at instant $t_{n+1}$ and define probabilities $y_j(t_n)$  of the
walker being at point $x_j$ at instant $t_n$.
Then, by
$$ y_k(t_{n+1}) = \sum_{j=-\infty}^{\infty}\, p_{j,k}\, u_j(t_n)\,,
 \q \sum_{k=-\infty}^{\infty}\, p_{j,k} =
 \sum_{j=-\infty}^{\infty}\, p_{j,k} =1 \,, \eqno(6.20)$$
with $	p_{j,k} = p_{k,j}\,,$
a symmetric random walk (more precisely a symmetric random jump)
model is described.
With  the approximation
$$  y_j(t_n) \approx \int_{(x_j-h/2)}^{(x_j+h/2)} \!\!\! u(x,t_n)\,dx
  \approx h\, u(x_j,t_n)\,,\eqno(6.21)$$
and introducing the "scaling parameter"
$$ \mu = {\tau \over h^\alpha} \, \rec{2\, |\cos (\alpha \pi/2)|}\,,
\eqno(6.22)$$ we have  solved
$$ {y_j(t_{n+1}) -y_j(t_n)\over \tau } =  -\,_hI^{-\alpha }\,y_j(t_n)\,,
   \eqno(6.23)$$
 for $y_j(t_{n+1})\,. $
 So we have proved to have a consistent
(for $\, h\to 0$) symmetric random walk approximation  to (6.18)
by taking
\pni
{\it i)} for $0<\alpha <1\,, $	 $\; 0<\mu \le 1/2\,,$
$$
\null \cases{
  _hI^{-\alpha }\,y_j(t_n) = \mu \, {\ds {h^\alpha \over \tau}}
  \,\l[_hI_+^{-\alpha }\,y_j(t_n) + \, _hI_-^{-\alpha }\,y_j(t_n)\r]\,,
   &\null \cr\cr
  p_{j,j}   = 1-2\mu \,, \;
  p_{j,j\pm k} = \mu \,\l\vert \alphak \r\vert\,, \q
   k \ge 1\,; & \null \cr}
 \eqno(6.24)  $$
{\it ii)} for $1<\alpha \le 2\,, $   $\; 0<\mu \le 1/(2\alpha)\,,$
$$ \null \cases{
   _hI^{-\alpha }\,y_j(t_n) = \mu \, {\ds {h^\alpha \over \tau}}
   \,\l[_hI_+^{-\alpha}\,y_{j+1}(t_n)+
    \,_hI_-^{-\alpha}\,y_{j-1}(t_n)\r]\,, \cr\cr
    p_{j,j} = 1-2\mu\,\alpha  \,, \;
    p_{j,j \pm 1} = \mu \,\l[1 +\alphatwo\r] \,, \; \cr\cr
    p_{j,j \pm k} = \mu \,\l\vert \alphakk \r\vert\,, \q
   k \ge 2\,. &\null \cr}
      \eqno(6.25) $$
We note that our random walk model is not only symmetric, but also
homogeneous, the transition probabilities $p_{j,j \pm k} $ 
 not depending on the index $j\,. $


\vsp
In the special case
 $\alpha =2$ we recover from (6.25) the well-known
three-point approximation of   the heat equation, because
$p_{j,j\pm k} =0$ for
 $k\ge 2\,.$
This means that for approximation of common diffusion
 only jumps of one step to the right or one to the left
or jumps of width zero occur, whereas
for
$0<\alpha <2$  ($\alpha \ne 1$) arbitrary large jumps
occur with power-like decaying probability, as it turns out
from the asymptotic analysis for the transition probabilities
given in (6.24-25).
In fact,  as $k \to \infty\,,$
one finds 
 $$ p_{j , j+k}
 \sim {(\tau /h^\alpha ) \over \pi}\,
 \Gamma(\alpha +1) \, \sin \l(\alpha{\pi\over 2}\r) \, k^{-(\alpha+1)}
 \,,\q 0<\alpha <2\,. \eqno(6.26) $$
This result thus  provides the discrete counterpart
of the	asymptotic behaviour
of the long
 power-law tails of the symmetric stable  distributions,
as  foreseen by (6.13) when $0< \alpha <2\,. $


%% file: mainardi_BUDAPEST97_7.tex

\section{Conclusions}

   We have treated two generalizations of the standard,
one-dimensional, diffusion equation, namely,
the {\it time-fractional} diffusion equation
and  the {\it symmetric space-fractional} diffusion equation.
For these equations we have derived the fundamental solutions 
using the transform methods of	Fourier and Laplace, and
exhibited their connections to {\it extremal} and {\it symmetric}
stable probability densities, evolving on time	 or variable in space.
For the {\it symmetric space-fractional} diffusion equation we have
presented a stationary (in time), homogeneous (in
space)	symmetric random walk model, discrete in space and time, the
step-lengths of the spatial grid and the time lapses between transitions
properly scaled. In the limit of infinitesimally fine discretization this
model (based on the {\it Gr\"unwald-Letnikov} approximation to fractional
derivatives) is consistent with the continuous diffusion process, \ie
convergent if interpreted as a difference scheme in the sense of
numerical analysis\footnote{
Further generalizations 
have been considered by us and our collaborators in other papers,
in which we have given a derivation of discrete
random walk models related to  more general
{\it space-time fractional diffusion equations}. For a comprehensive
analysis, see Gorenflo et al. (2002). Readers interested to the fundamental
solutions of these  fractional  diffusion equations
are referred to the paper by Mainardi et al. (2001) where
analytical expressions and numerical plots are found.}.
\vsp
    From the mathematical viewpoint the field of such "fractional"
generalizations is fascinating	as there several mathematical disciplines
meet and come to a fruitful interplay:
\eg probability theory and stochastic processes,
integro-differential equations,
transform theory, special functions, numerical analysis.
As one may take from our References, 
one can  observe that since some decades there is an ever growing interest
in using the concepts of fractional calculus among physicists and
economists.
Among  economists  we like to refer the reader
to a collection of papers on  the topic of "Fractional Differencing and Long
Memory Processes", edited by Baillie \& King (1996).

%% file: mainardi_BUDAPEST97_A.tex
\noindent
\section*{Appendix A: The Riemann-Liouville Fractional Calculus}
Fractional calculus is the field of mathematical analysis which deals
with the investigation and applications of integrals and
derivatives of arbitrary order.
The term {\it fractional} is a misnomer, but it is
retained 
following the  prevailing use.
This appendix is  mostly based on the recent review by Gorenflo \&
Mainardi  (1997).
For more details on the classical treatment of	fractional
calculus the reader is referred to Erd\'elyi (1954), Oldham \& Spanier
(1974), Samko {\it et al.} (1987-1993) and  Miller \& Ross (1993).
\vsp
According to the Riemann-Liouville approach  to fractional calculus,
the notion of fractional Integral of order $\alpha$
($\alpha >0$)
 is a natural consequence
of the well known formula (usually attributed to Cauchy),
that reduces the calculation of the $n-$fold primitive of a function
$f(t)$ to a single integral of convolution type.
In our notation the Cauchy formula reads
$$
    J^n f(t) := f_{n}(t)=
 \rec{(n-1)!}\, \int_0^t \!\!  (t-\tau )^{n-1}\,f(\tau) \, d\tau\,,
    \q t > 0\,, \q n \in \NN\,,  \eqno(A.1) $$
where $\NN$ is the set of positive integers.
From this definition we note that $f_{n}(t)$
vanishes at $t=0$ with its
derivatives of order $1,2, \dots, n-1\,. $
For convention we require that	$f(t)$ and henceforth
$f_{n}(t)$ be a {\it causal} function, \ie identically
vanishing for $t<0$.
In a natural way  one is  led
to extend the above formula
from positive integer values of the index to any positive real values
by using the Gamma function.
Indeed, noting that $(n-1)!= \Gamma(n)\,, $
and introducing the arbitrary {\it positive} real number
 $\alpha\,, $
one defines  the
 \ \underbar{{\it Fractional Integral of order} $\alpha >0\,:$}
$$
J^\alpha \,f(t) :=
      \rec{\Gamma(\alpha )}\,
 \int_0^t (t-\tau )^{\alpha -1}\, f(\tau )\,d\tau \,,
   \q t > 0\,,\q \alpha  \in \RR^+
 \,,  \eqno(A.2) $$
where $\RR^+$ is the set of positive real numbers.
For complementation we define
$J^0 := \II\, $ ({Identity operator)}, \ie we mean
$J^0\, f(t) = f(t)\,. $ Furthermore,
by $J^\alpha f(0^+)$ we mean the limit (if it exists)
of $J^\alpha f(t)$ for $t\to 0^+\,;$ this limit may be infinite.
\newpage
\vsp
We note the {\it semigroup property}
$J^\alpha J^\beta = J^{\alpha +\beta}\,,
   \; \alpha\,,\;\beta	\ge 0\,,$
which implies the {\it commutative property}
$J^\beta  J^\alpha= J^\alpha J^\beta\,,$
and  the effect of our operators $J^\alpha$
on the power functions
$$
J^\alpha t^\gamma ={\Gamma (\gamma +1)\over \Gamma(\gamma +1+\alpha)}\,
		   t^{\gamma+\alpha}\,, \q \alpha \ge 0\,,
  \q \gamma >-1\,, \q t>0\,.
\eqno (A.3)
$$
These properties  are of course a natural generalization
of those known when the order is a positive integer.
\vsp
Introducing
the Laplace transform by the notation
$ {\cal{L}}\, \l\{  f(t) \r\}  := \int_0^\infty \!\!
   \e^{-st}\, f(t)\, dt = \widetilde f(s)\,, \; s \in \CC\,,$
and  using the sign $\div$ to denote a Laplace transform pair,
\ie
$ f(t) \div  \widetilde f(s) \,, $
we note the following rule for the   Laplace transform of
the fractional integral,
$$	   J^\alpha \,f(t) \div
     {\widetilde f(s)\over s^\alpha}\,,\q \alpha \ge 0\,,  \eqno(A.4)$$
which is the generalization
of the case with an $n$-fold repeated integral. 
\vsp
After the notion of fractional integral,
that of fractional derivative of order $\alpha$
($\alpha >0$)
becomes a natural requirement and one is attempted to
substitute $\alpha $ with $-\alpha $ in the above formulas.
However, this generalization  needs some care  in order to
guarantee the convergence of  the integrals   and
preserve the
well known properties of the ordinary derivative of integer
order.
\vsp
 Denoting by $D^n\,$ with $ n\in \NN\,, $
the operator of the derivative of order $n\,,$	we first note that
$ D^n \, J^n = \II\,, \;   J^n \, D^n \ne \II\,,\q n\in \NN \,,
$
\ie $D^n$ is left-inverse (and not right-inverse) to
the corresponding integral operator $J^n\,. $
In fact we easily recognize from (A.1) that
$$  J^n \, D^n \, f(t) = f(t) - \sum_{k=0}^{n-1}
	f^{(k)}(0^+) \, {t^k\over k!}\,, \q t>0\,. \eqno(A.5)$$
\vsp
As a consequence we expect that $D^\alpha $ is defined as left-inverse
to $J^\alpha $.  For this purpose, introducing the positive
integer $m$ such that $m-1 <\alpha \le m\,, $
one defines the
 \underbar{{\it Fractional Derivative of order} $\alpha >0 $} :
$$
D^\alpha \,f(t) :=
 D^m \, J^{m-\alpha} \, f(t)\,, \quad  m-1<\alpha \le m\,, \quad m \in \NN\,, 
  \eqno(A.6) $$
namely
$$D^\alpha \,f(t) \!=\! 
\cases{
{\ds{d^m\over dt^m}\left[
   \rec{\Gamma(m-\alpha)}\int_0^t \!\!
    {f(\tau )\over (t-\tau )^{\alpha +1-m}} \,d\tau\right]} ,& $\!\! m-1<\alpha <m$,\cr
{\ds {d^m\over dt^m}\, f(t)}\,, & $\!\! \alpha =m.$\cr}
\eqno(A.6')$$	
Defining for complementation $D^0 = J^0 =\II\,, $ then
we easily recognize that
$ D^\alpha \, J^\alpha = \II \,,$  $\, \alpha \ge 0\,,$
and
$$ D^{\alpha}\, t^{\gamma}=
   {\Gamma(\gamma +1)\over\Gamma(\gamma +1-\alpha)}\,
     t^{\gamma-\alpha}\,,
 \q \alpha \ge 0\,,
  \q \gamma >-1\,, \q t>0\,.
\eqno (A.7)
$$
Of course, these properties are a natural generalization
of those known when the order is a positive integer.
\vsp
Note the remarkable fact that the fractional derivative $D^\alpha\, f$
is not	zero
for the constant function $f(t)\equiv 1$ if $\alpha \not \in {\NN}\,. $
In fact, (A.7) with $\gamma =0$ teaches us that
$$
D^\alpha 1 = {t^{-\alpha}\over \Gamma(1-\alpha)}\,,\q \alpha\ge 0\,,
\q t>0\,.  \eqno (A.8)
$$
This, of course, is $\equiv 0$ for $\alpha \in{\NN}$, due to the
poles of the gamma function in the points $0,-1,-2,\dots$.
We now observe that an alternative definition
of fractional derivative, originally introduced by Caputo (1967)
(1969)
in the late sixties and
adopted by Caputo and Mainardi (1971)
in the framework  of the theory of {\it Linear Viscoelasticity},
is
$$
D_*^\alpha  \, f(t) :=
  J^{m-\alpha}\, D^{m} \, f(t) 
 \quad m-1<\alpha \le m\,, \quad m\in \NN\,,
  \eqno(A.9) $$
 namely
$$D*^\alpha \,f(t) = 
\cases{
   {\ds \rec{\Gamma(m-\alpha)}\,\int_0^t \!\!
    {f^{(m)}(\tau )\over (t-\tau )^{\alpha +1-m}} \,d\tau} \,,& $m-1<\alpha <m$,\cr
{\ds {d^m\over dt^m}\, f(t)}\,, & $\alpha =m$.\cr
}
 \eqno(A.9')$$	 
This definition is of course more restrictive than (A.6), in that
requires the absolute integrability of the  derivative of order $m$.
Whenever we use the operator   $D_*^\alpha$   we (tacitly) assume that
     this condition is met.
 We  easily recognize that in general
$$  D^\alpha\, f(t) := D^{m} \, J^{m-\alpha} \, f(t)
 \ne J^{m-\alpha}\, D^{m} \, f(t):= D_*^\alpha \, f(t)\,,
 \eqno(A.10)
 $$
 unless   the function	$f(t)$ along with its first $m-1$ derivatives
 vanishes at $t=0^+$.
In fact, assuming that
the passage of the $m$-derivative under
the integral is legitimate, one     
recognizes that,  for $ m-1 <\alpha  < m \,$  and $t>0\,, $
$$
    D^\alpha \, f(t) =
  D_*^\alpha   \, f(t) +
  \sum_{k=0}^{m-1}   {t^{k-\alpha}\over\Gamma(k-\alpha +1)}
    \, f^{(k)}(0^+) \,, \eqno(A.11)    $$
 and therefore, recalling the fractional derivative of the power
functions (A.7),
$$
   D^\alpha \l( f(t) -
 \sum_{k=0}^{m-1} {t^k \over k!} \, f^{(k)} (0^+)\r)
     =	D_*^\alpha  \, f(t)  \,.\eqno(A.12)  $$
The alternative definition (A.9) for the
fractional derivative  thus incorporates the initial values
of the function and of its integer derivatives of lower order.
The subtraction of the Taylor polynomial of degree $m-1$ at $t=0^+$
from $f(t)$ means  a sort of
regularization	of the fractional derivative.
In particular, according to this definition,
the relevant property for which the fractional derivative
of a constant is still zero can be easily recognized,
 \ie
$$ D_*^\alpha  1 \equiv 0\,,\q	 \alpha >0\,.\eqno(A.13)$$
We now explore the most relevant differences between the two
fractional derivatives (A.6) and (A.9). We agree to
denote (A.9) as the {\it Caputo fractional derivative}
to distinguish it from the standard Riemann-Liouville fractional
derivative (A.6).
We observe, again by looking at (A.7), that
$D^\alpha t^{\alpha -1} \equiv 0\,,$ $\alpha>0\,, \; t>0\,.$
\vsp
From above  we thus recognize
the following statements about functions
which  for $t>0\, $   admit the same fractional derivative
of    order $\alpha \,, $
with $m-1 <\alpha \le m\,,$ $\; m \in \NN\,, $
$$    D^\alpha \, f(t) = D^\alpha  \, g(t)
   \,  \Longleftrightarrow  \,
  f(t) = g(t) + \sum_{j=1}^m c_j\, t^{\alpha-j} \,,
    \eqno(A.14) $$
$$    D_* ^\alpha \, f(t) = D_*^\alpha	\, g(t)
   \,  \Longleftrightarrow  \,
  f(t) = g(t) +  \sum_{j=1}^m c_j\, t^{m-j} \,.
    \eqno(A.15) $$
In these formulas the coefficients $c_j$ are arbitrary constants.
\vsp
For the two definitions we also note a difference
with respect to the {\it formal} $\,$ limit  as
 $\alpha \to {(m-1)}^+\,;$ from (A.6) and (A.9) we obtain
respectively,
$$ D^\alpha \,f(t) \to	  D^m\,  J\, f(t) = D^{m-1}\, f(t)
   \,; \eqno(A.16) $$
$$
 D_*^\alpha \, f(t) \to J\, D^m\, f(t) =
       D^{m-1}\, f(t) - f^{(m-1)} (0^+)\,. \eqno(A.17) $$
\vsp
We now consider the {\it Laplace transform} of the two fractional
derivatives.
For the standard fractional derivative $D^\alpha $
the Laplace transform,	assumed to exist,  requires the knowledge of the
(bounded) initial values of the fractional integral $J^{m-\alpha }$
and of its integer  derivatives of order $k =1,2, \dots, m-1\,. $
The corresponding rule reads, in our notation,
$$ D^\alpha \, f(t) \div
      s^\alpha\,  \widetilde f(s)
   -\sum_{k=0}^{m-1}  D^k\, J^{(m-\alpha)}\,f(0^+) \, s^{m -1-k}\,,
   \eqno(A.18)$$
where $ m-1<\alpha \le m \,.$
\vsp
The {\it Caputo fractional derivative} appears more suitable to
be treated by the Laplace transform technique in that it requires
the knowledge of the (bounded)
initial values of the function
and of its integer  derivatives of order $k =1,2, \dots, m-1\,, $
in analogy with the case when $\alpha =m\,. $
In fact,
by using (A.4) and noting that
$$  
J^\alpha  \, D_*^\alpha \, f(t)= 
    J^\alpha\, J^{m-\alpha }\, D^m \, f(t) =
     J^m\, D^m \, f(t) 
  = f(t) -	{\ds\sum_{k=0}^{m-1}} {f^{(k)}(0^+)}\, {\ds{t^k \over k!}}
  \,,	 
\eqno(A.19)$$
we easily prove  the following rule for the Laplace transform,
$$ D_*^\alpha \, f(t) \div
      s^\alpha\,  \widetilde f(s)
   -\sum_{k=0}^{m-1}  f^{(k)}(0^+) \, s^{\alpha -1-k}\,,
  \q m-1<\alpha \le m \,. \eqno(A.20)$$
Indeed, the  result (A.20), first stated by Caputo (1969) by using the
Fubini-Tonelli theorem, appears  as the most "natural"
generalization of the corresponding result well known for $\alpha =m\,. $
\vsp
Gorenflo and Mainardi (1997)
have pointed out the major utility of the
Caputo fractional derivative
in the treatment of differential equations of fractional
order for {\it physical applications}.
In fact, in physical problems,	the initial conditions are usually
expressed in terms of a given number of bounded values assumed by the
field variable and its derivatives of integer order,
no matter if
the governing evolution equation may be a generic integro-differential
equation and therefore, in particular,	a  fractional differential
equation\footnote{
 We note that the {\it Caputo fractional  derivative} 
 was so named after the book by Podlubny (1999). It coincides with that
 introduced,  independently and  a few later,
 by  Dzherbashyan and  Nersesyan (1968) 
 as a regularization of the Riemann-Liouville fractional derivative.
 Nowadays, some Authors refer to it as the {\it Caputo-Dzherbashyan fractional derivative}. 
The prominent role of this fractional derivative in treating initial value problems
  was   recognized   in interesting papers by  Kochubei (1989), (1990).}.
\vsp
We now analyze
the most simple differential equations of fractional order,
including those
which, by means of  fractional derivatives, generalize the
well-known ordinary differential equations related to
relaxation and oscillation phenomena.
Generally speaking, we consider
the following differential equation of fractional order
$\alpha >0\,, $
$$ D_*^\alpha \, u(t) =
D^\alpha  \l( u(t) - \sum_{k=0}^{m-1} {t^k \over k!} \,
u^{(k)} (0^+)\r)    = - u(t) + q(t) \,,  \q t>0\,, \eqno(A.21)$$
where $u=u(t)$ is the field variable and $q(t)$ is a given function.
Here $m$ is a positive integer uniquely defined
by $m-1 <\alpha  \le m\,, $
which provides the number of the prescribed initial values
$u^{(k)}(0^+) = c_k\,, \; k=0,1,2, \dots , m-1\,. $
 Implicit in the form of (A.21) is our desire to obtain solutions
 $u(t)$ for which the $u^{(k)}(t)$ are continuous.
In particular, the cases of {\it fractional relaxation} and
{\it fractional oscillation}
are obtained for $0<\alpha <1$ and $ 1<\alpha <2\,, $	respectively
\vsp
The application of the Laplace transform
through  the Caputo formula (A.20)
yields
$$ \bar u(s) = \sum_{k=0}^{m-1} c_k \,{s^{\alpha -k -1} \over s^\alpha +1}
       + \rec{s^\alpha+1 }\, \bar q(s)\,.  \eqno(A.22)$$
\vsp
Now, in order to obtain the Laplace inversion of (A.22), we need to
recall the {\it Mittag-Leffler} function of order $\alpha >0\,, $
$E_\alpha(z)\,. $
This function, so named from the great Swedish mathematician
who introduced it at the beginning of
this century,
is defined by the following series and integral representation,
valid in the whole complex plane,
$$
E_\alpha (z) =
   \sum_{n=0}^\infty {z^n\over \Gamma (\alpha n+1)} =
 \rec{2\pi i}\,  \int_{Ha} {\sigma ^{\alpha -1}
 \, \e^{\,\sigma } \over \sigma ^\alpha -z}\,
	d\sigma  \,, \q \alpha >0\,. \eqno(A.23)$$
Here  ${Ha}\,$	denotes the {Hankel path}, i.e.
a loop which
starts and ends at $-\infty$ and encircles the circular disk
$|\sigma | \le |z|^{1/\alpha }$ in the positive sense.
It turns out that
$E_\alpha (z) $ is an {entire function}
 of order $\rho =1/\alpha \,$ and type $1\,. $
\vsp
The  {\it Mittag-Leffler} function provides a simple generalization of the
  exponential function, to which it reduces for $\alpha = 1\,. $
Particular cases  from which elementary functions
are   recovered,  
are
$$ E_2\l(+z^2\r) =  \cosh \, z\,, \qq
   E_2\l(-z^2\r) =  \cos \, z\,, \qq z\in \CC \,,   \eqno(A.24)$$
and
$$ E_{1/2} (\pm  z^{1/2}) =
     \e^{\ds z} \, \l[ 1+{\erf}\, (\pm z^{1/2})\r ] =
   \e^{\ds z} \, {\erfc} \, (\mp z^{1/2})\,,\q z\in \CC \,,
 \eqno(A.25)$$
where $\erf$ ($\erfc$) denotes the (complementary) error function.
 defined as
 $$  {\erf} \,(z) := {2\over \sqrt{\pi}}\,\int_0^z \e^{\ds -u^2}\,du
 \,,  \q   {\erfc} \,(z) :=  1 - {\erf}\, (z)\,,
  \q z\in \CC\,. $$
A noteworthy property of the {\it Mittag-Leffler} function
is based on the following {\it duplication formula}
$$ E_\alpha (z) = \rec{2}\,
 \l[ E_{\alpha /2}(+z^{1/2})+	E_{\alpha /2}(-z^{1/2})\r]\,.
   \eqno(A.26) $$
In (A.25-26) we agree to denote by $z^{1/2}$ the main branch
of the complex root of $z\,. $
\vsp
     The {\it Mittag-Leffler} function	is connected to the
Laplace integral through the equation
$$ {
\int_0^\infty \!\!\! \e^{-u} \, E_\alpha \l(u^\alpha \,z\r) \, du
   = \rec{1-z}
  \q \alpha   >  0
   }\,.\eqno(A.27)$$
The integral at the L.H.S. was evaluated by Mittag-Leffler who showed that
the region of its convergence contains the
unit circle and is bounded by the line
${\rm Re}\, z^{1/\alpha } =1\,. $
The above integral is fundamental in the  evaluation of
the {Laplace transform} of 
$E_\alpha \l(-\lambda \, t^\alpha \r)$
with $\alpha   >0$ and $\lambda  \in \CC\,. $
In fact, putting  in (A.27)
$u=st$ and $u^\alpha \, z = -\lambda \, t^\alpha $  with
$t \ge 0\, $ and $\lambda \in \CC \,, $
we get the Laplace
transform pair
$$
 E_\alpha \l(-\lambda \, t^\alpha \r)
   \div {s^{\alpha -1}\over s^\alpha +\lambda} \,,
 \q Re \, s > |\lambda |^{1/\alpha }
\,. \eqno(A.28)$$
Then, using (A.28), we put for	$k=0,1,\dots,m-1\,,$
$$ u_k(t) := J^k e_\alpha (t) \, \div \,
   {s^{\alpha -k -1} \over s^\alpha +1}  \,,\q
    e_\alpha (t)  :=E_\alpha(-t^\alpha)\,,
  \eqno(A.29)$$
and,  from inversion of the Laplace transforms in (A.22),
we find
$$ u(t) = \sum_{k=0}^{m-1} c_k \, u_k (t)
   - \int_0^t \!\! q(t-\tau )\, u_0 ^\prime(\tau)\, d\tau \,.
   \eqno (A.30)$$
In particular, the  formula (A.30)
encompasses the solutions
for $\alpha = 1\,,\,2\,,$  since
$e_1(t) = {\exp} (-t)\,,$ $\, e_2(t) = \cos \, t\,. $
When $\alpha$ is not integer, namely for $m-1 <\alpha <m\,, $
we note that $m-1$ represents
the integer part of $\alpha$ (usually denoted by $[\alpha]$)
and $m$  the number  of initial  conditions
necessary and sufficient to ensure the uniqueness of the
solution $u(t)$. Thus the $m$ functions
$u_k(t) = J^k e_\alpha (t)$ with $k =0, 1, \dots, m-1 \, $
represent those particular solutions
of the {\it homogeneous} equation which satisfy
the initial conditions
$ u_k^{(h)} (0^+)  = \delta _{k\,h} \,,$
$\; h,k =0, 1, \dots, m-1 \, ,$
and therefore they represent the {\it fundamental solutions}
of the fractional equation (A.21), in analogy with the case
$\alpha =m\,. $  Furthermore, the function
$ u_\delta (t) = -u_0^\prime(t) =  -e_\alpha ^\prime (t)$
represents the {\it impulse-response solution}.
\vsp
The {\it Mittag-Leffler} function of order less than one turns out to be
related through the Laplace integral
to another special function of Wright type, denoted by
$M(z,\nu) $ with $0<\nu <1\,, $ following the notation
introduced by Mainardi (1994, 1995).
Since this function turns out to be relevant in the general framework
of fractional calculus with special regard to stable probability
distributions, we are going to summarize its basing properties.
For more details on this function, see Mainardi (1997), Appendix A.
\vsp
Let us first recall the more general {\it Wright} function
$W_{\lambda ,\mu }(z)\,, z\in \CC\,, $ with $\lambda >-1$ and $\mu >0\,. $
This function, so named from the British mathematician
who introduced it between  1933 and 1941,
is defined by the following series and integral representation,
valid in the whole complex plane,
  $$ W_{\lambda ,\mu }(z ) =
   \sum_{n=0}^{\infty}{z^n\over n!\, \Gamma(\lambda  n + \mu )}=
   {1\over 2\pi i}\,\int_{Ha}	\!\!
 \e^{\, \ds \sigma +z\sigma ^{-\lambda }} \,
   {d\sigma \over \sigma^{\mu}} \,,
   \eqno(A.31)$$
where $Ha$ denotes the Hankel path.
It is possible to prove that the {\it Wright} function is entire of order
$1/(1+\lambda)\,, $ hence
of exponential type if $\lambda \ge 0\,. $
The case $\lambda =0$ is trivial since
$  W_{0, \mu }(z) = { \e^{\, z}/ \Gamma(\mu )}\,.$
The case $\lambda = -\nu  \,, \,\mu =1-\nu   $ with $0<\nu   <1$
provides the function $M(z,\nu	 )$ of special interest for us.
Specifically, we have
 $$ M(z;\nu  ) :=  W_{-\nu  , 1-\nu  }(-z)= \rec{\nu   \,z}\,
     W_{-\nu  , 0}(-z)	\,, \q 0<\nu  <1\,, \eqno(A.32)$$
and therefore from (A.31-32)
   $$ \begin{array}{lll}
M(z;\nu) &= 
  {\ds\rec{\pi}}\,
  {\ds\sum_{n=1}^{\infty}}\,{\ds{(-z)^{n-1} \over (n-1)!}}\,
   \Gamma(\nu \,n)  \,\sin \, (\nu\, n\,\pi)
 \\  
\null  &= {\ds{1\over 2\pi i}}\,{\ds \int_{Ha}} \!\!
 \e^{\ds \,\sigma -z\sigma^\nu}\,  {\ds{d\sigma\over \sigma^{1-\nu}}} \,,
 \q  0<\nu  <1\,.
   \end{array}	    \eqno(A.33)$$
In the series representation we have used
the  reflection formula for the Gamma function,
 $ \Gamma(x)\,\Gamma(1-x)  =\pi /\sin\,\pi x\,.$
Explicit expressions of $M(z;\nu)$ in terms of simpler known
functions are expected in particular cases when $\nu$ is a rational
number.  Relevant cases
are  $\nu   =1/2\,, \,1/3$ for which


$$ M(z;1/2)   = \rec{\sqrt{\pi}}\, \exp \l(-{\,z^2/ 4}\r)\,,\eqno(A.34)$$
$$   M(z;1/3) =  3^{2/3} \, {\Ai} \l( {z/ 3^{1/3}}\r) \,,
 \eqno(A.35)$$
where $\, {\Ai}\,$ denotes the {\it Airy} function.
\vsp
When the argument is real and positive, \ie $ z =r >0\,, $
the existence of the Laplace transform	    of $M(r;\nu  )$
is ensured by the asymptotic behaviour, as derived by Mainardi \&
Tomirotti (1995), as $ r\to +\infty\,,$
$$ M(r/\nu   ;\nu   ) \sim
   a(\nu   )\, r^{\ds{(\nu   -1/2)/(1-\nu  )}}	\;
   \exp \, \l[- b(\nu  )\,r^{\ds {1/(1-\nu  )}}\r]\,,
 \eqno(A.36)$$
where $ a(\nu  ) = {1/	\sqrt{2\pi\,(1-\nu  )}}\,,  \;
  b(\nu  ) = {(1-\nu  )/ \nu  }  \,.$
\vsp
It is an instructive exercise
to derive the Laplace transform by interchanging
the Laplace integral with the Hankel integral in (A.33) and
recalling  the integral representation (A.23) of the Mittag-Leffler
function.
We obtain the Laplace transform pair
$$  M(r;\nu   ) \,\div\,  E_\nu   (-s)\,, \q 0<\nu  <1 \,.
  \eqno(A.37)$$
For $\nu   =1/2\,, $ (A.37) with (A.25) and (A.34) provides  the
result, see \eg Doetsch (1974),
$$  M(r;1/2 ) := \rec{\sqrt{\pi}}\, \exp \l(-{\,r^2/ 4}\r)
\,\div\,  E_{1/2} (-s) := \exp \l(s^2\r)\, {\erfc}\l( s\r)\,.
   \eqno(A.38) $$
 It would be noted  that, since  $M(r,\nu   )$	is not of
exponential order,
transforming term-by-term the Taylor series of
$M(r;\nu  )\, $  yields  a series
of negative powers of $s\,, $ which represents the asymptotic
expansion
of  $E_\nu   (-s)$ as $s\to \infty\,$ in a certain sector around the
real axis.
\vsp
We also note that (A.37) with (A.23) allows us to compute
the moments of any real order $ \delta	\ge 0\,$ of $M(r;\nu)$ in the
positive real axis. We obtain
$$\int_0^{+\infty} \!\!\! r^{\,\ds \delta}\, M(r;\nu)\, dr
  = {\Gamma(\delta +1)\over \Gamma(\nu	\delta	+1)}\,,
   \q \delta  \ge 0\,.
\eqno(A.39)$$
 When $\delta$ is integer we note that the moments
are provided by the derivatives of the Mittag-Leffler function
in the origin, \ie
  $$\int_0^{+\infty} \!\!\! r^{\, \ds n}\, M(r;\nu )\, dr
     = \lim_{s\to 0} \, (-1)^n \, {d^n\over ds^n}\, E_\nu (-s)
  = {\Gamma(n+1)\over \Gamma(\nu   n+1)}\,,
\eqno(A.40)$$
where $n =0,1,2,\dots\,. $
The normalization condition
$\int_0^{\infty} M(r;\nu   )\, dr =E_\nu   (0) = 1\,$
is recovered for $ n=0\,. $
The relation with the Mittag-Leffler function
stated in (A.40) can be extended  to the moments of non integer order
if we replace
the ordinary derivative, of order $n$, with the corresponding fractional
derivative, of order $\delta \ne n$,  in the {\it Caputo} sense.


\vsp
Another exercise on the function $M$ concerns  the inversion of
the Laplace transform
$\exp (-s^\nu  )\,, $ either by  the complex integral
formula or  by	the formal series method.
We obtain  the	Laplace transform pair
$$
   {\nu   \over r^{\nu	 +1}}\,  M\l( 1/{r^\nu	 };\nu	 \r)\,\div\,
    \exp \l(\ds -s^\nu	\r)\,, \q  0<\nu   <1\,. \eqno(A.41)$$
For $\nu   =1/2\,, $ (A.41) with (A.34) provides  the known result, see \eg
Doetsch (1974),
$$ {1\over 2\, r^{3/2}}\,  M(1/{r}^{1/2};1/2) :=
 \rec{2\sqrt{\pi}\, r^{3/2}}\, \exp \l[-{\,1/(4r)}\r] \,\div\,
  \exp \l(-s^{1/2}\r)\,.
   \eqno(A.42) $$
  We recall that a rigorous proof of (A.41) was formerly given by
Pollard (1946), based on a formal result by Humbert (1945).
The Laplace transform pair was also obtained  by
Mikusi\'nski (1959) and, albeit unaware of  the previous results,
by  Buchen \& Mainardi (1975) in a formal way.


%% file: mainardi_BUDAPEST97_B.tex

\noindent
\section*{Appendix B: The Stable Probability Distributions}
 The stable distributions are a fascinating
and fruitful area of research in probability theory;
furthermore, nowadays,	they provide valuable models
in physics, astronomy, economics, and communication theory.
\vsp
The general class of stable distributions
was introduced and given this name by the French mathematician
Paul L\'evy
in the early 1920's, see L\'evy (1924, 1925).
The inspiration  for  L\'evy
was the desire to generalize the celebrated {\it Central Limit Theorem},
according to which any probability distribution with finite variance
belongs to the domain of attraction of the Gaussian distribution.
\vsp
Formerly, the topic
attracted only moderate attention from the leading experts,
though there were also enthusiasts, of whom  the Russian
mathematician Alexander Yakovlevich
Khintchine should be mentioned first of all.
The concept  of stable distributions took full shape
in 1937 with the appearance of L\'evy's  monograph, 
see L\'evy (1937-1954),
soon followed by Khintchine's monograph, see Khintchine (1938).
\vsp
The theory and properties of stable
distributions are discussed
in some classical books on probability theory including
Gnedenko \& Kolmogorov (1949-1954), Lukacs (1960-1970),
Feller (1966-1971), Breiman (1968-1992),
Chung (1968-1974) and  Laha \& Rohatgi	(1979).
Also treatises on fractals devote particular attention to stable
distributions  in view of their properties of scale invariance,
see  \eg Mandelbrot (1982) and	Takayasu (1990).
Sets of tables and graphs have been
provided by Mandelbrot \& Zarnfaller (1959), Fama \& Roll (1968),
Bo'lshev \& Al. (1968) and Holt \& Crow (1973).
\vsp
 Only recently,	 monographs  devoted solely to stable
distributions and related stochastic processes
have been appeared, \ie
   Zolotarev (1983-1986),
  Janicki \& Weron (1994),
  Samorodnitsky \& Taqqu (1994),
  Uchaikin \& Zolotarev  (1999).
We now can  cite the paper by Mainardi, Luchko \& Pagnini (2001)
where the reader can find (convergent and  asymptotic) representations
and plots of the symmetric and non-symmetric stable densities  generated by 
fractional diffusion equations.  
\vsp
Stable distributions have three {\it exclusive} properties,
which can
be briefly summarized stating that they 
1) are {\it invariant under addition}, 
2) possess their {\it own domain of attraction}, 
and 3) admit a {\it canonic characteristic function}.
\vsp
Let us now illustrate the above properties which, providing
necessary and sufficient conditions, can be assumed as
equivalent definitions for a stable distribution.
We  recall the basic results without proof.
\vsp
{\it A random variable $X$ is said to have a stable distribution}
$P(x) = {\Prob} \, \{X \le x\}$  
{\it if for
any $n\ge 2\,, $ there is a positive number $c_n$ and a real number
$d_n$ such that
$$ X_1 +X_2 +\dots + X_n \,{\mathop=^{d}}\, c_n\, X +d_n\,, \eqno(B.1)$$
where $X_1, X_2, \dots X_n$ denote mutually independent random variables
with  common distribution  $P(x)$ with $X\,. $}
Here the   notation $\,\ds {\mathop=^{d}}\,$ denotes equality
in distribution, \ie  means that
the random variables on both sides have the same probability distribution.
\vsp
When mutually independent random variables have a common distribution
[shared with a given random variable $X$], we  also refer to them as
independent, identically distributed (i.i.d) random variables
[independent copies of $X$].
In general, the sum of i.i.d. random variables becomes a random variable
with a distribution of different form. However, for  independent
random variables   with a common {\it stable} distribution,
the sum obeys to a distribution of the same type, which differs
from the original one only for a scaling ($c_n$)  and possibly
for a shift ($d_n$).
When in (B.1) the $d_n=0$ the distribution is called
 {\it strictly	 stable}.
  \vsp
 It is known, see Feller (1966-1971), that the norming constants 
in (B.1) are of the form
$$c_n= n^{1/\alpha} \q {\rm with} \q 0<\alpha \le 2\,.\eqno(B.2)$$
The parameter $\alpha $ is called   the {\it characteristic exponent}
or the {\it index of stability} of the stable distribution.
\vsp
We agree to use the notation $X \sim P_\alpha (x)$ to denote
that the random variable $X$ has a stable probability distribution
with characteristic exponent $\alpha \,.$
 We simply refer to $P(x)\,, $
$p(x) := dP/dx$ (probability density function = $pdf$)  and
$ X$ as 
{\it $\alpha$-stable distribution, density,  random variable},
respectively.
\vsp
 The  definition (B.1) with the theorem (B.2) can be stated
in an alternative version that needs only  two i.i.d. random variables.
see also Lukacs (1960-1970).  
{\it A random variable $X$ is said to have a stable distribution if
for any positive numbers $A$ and $B$, there is a positive number $C$
and a real number $D$ such that
$$ A \, X_1 + B \, X_2 \, {\mathop=^{d}}\,  C\, X +D \,, \eqno(B.3)$$
where $X_1$ and $X_2$ are independent copies of $X\,. $}
Then there is a number $\alpha \in (0,2]$ such that the number $C$
in (B.3) satisfies
$  C^\alpha =  A^\alpha + B^\alpha \,.$
\vsp
For a strictly stable distribution (B.3) holds with $D=0\,. $
This implies that all linear combinations of i.i.d. random variables
obeying to a strictly stable distribution is  a random variable
with the same type of distribution.
\vsp
A stable distribution is called {\it symmetric} if  the random variable
$-X$ has the same distribution. Of course, a {\it symmetric} stable
distribution is necessarily {\it  strictly stable}.
\vsp
Noteworthy examples of stable distributions
are provided by the Gaussian (or normal) law (with $\alpha =2$) and
by the
Cauchy-Lorentz law ($\alpha =1$).
The corresponding  $pdf$
 are known to be 
$$
p_G(x;\sigma, \mu) :=
{1\over \sqrt{2\pi}\,\sigma }\, \e^{\,\ds -(x-\mu)^2/(2 \sigma^2)}\,,
\q  x \in \RR\,,\eqno(B.4)$$
where $\sigma^2 $ denotes the variance and $\mu $ the mean, and
$$
p_C(x;\gamma ,\delta) :=
{1\over \pi}\,	 {\gamma   \over (x-\delta)^2+\gamma ^2}\,, \q
 x\in \RR\,,\eqno(B.5)$$
where $\gamma  $ denotes the semi-interquartile range
and $\delta $ the "shift".
 \vsp
Another (equivalent) definition states that stable distributions are the
only distributions that can  be obtained as limits of normalized sums of
i.i.d. random variables.
A random variable $X$ is said to have a
{\it domain of attraction},\ie if there is a sequence of
i.i.d. random variables $Y_1, Y_2, \dots $ and sequences of
positive numbers $\{\gamma _n\}$ and real numbers $\{\delta _n\}$, such
that $$ {Y_1 +Y_2 + \dots Y_n\over \gamma _n} + \delta _n
{\mathop\Rightarrow^{d}} X\,.
\eqno(B.6)$$
The notation $\,\ds {\mathop\Rightarrow^{d}}\, $  denotes
convergence in distribution.
\vsp
It is clear that the previous definition (B.1) yields (B.6), \eg,
by taking the $Y_i$s  to be independent and distributed like $X\,. $
The converse is easy to show, see Gnedenko \& Kolmogorov (1949-1954).
Therefore we can alternatively state that {\it a random variable
$X$ is said to have a stable distribution if it has
a domain of attraction.}
 \vsp
When $X$ is Gaussian and the $Y_i$s are i.i.d. with finite variance,
then (B.6) is the statement of the ordinary {\it Central Limit Theorem}.
The domain of attraction of $X$ is said {\it normal}
when $\gamma _n= n^{1/\alpha}\,; $  in general,
$\gamma _n = n^{1/\alpha}\, h(n)$ where $h(x)\,,\, x>0\,, $ is a slow
varying function at infinity, that is,
$\ds \lim_{x\to \infty} h(ux)/h(x) =1$ for all $u>0\,,$ see Feller (1971).
The function $h(x) = {\log}\, x\,, $ for example, is slowly varying
at infinity.
\vsp
Another definition specifies the {\it canonic form} that the
{\it characteristic function} ($cf$) of a stable distribution of
index $\alpha $ must have. Recalling that the $cf$ is the Fourier
transform of the $pdf$, we use the notation
$ \hat p_\alpha (\kappa ) := \,
   \langle  {\exp} \, ({i\kappa X})\rangle \, \div \, p_\alpha (x)\,. $
We first note that a stable distribution
is also {\it infinitely divisible}, \ie for every positive integer
$n$ its  $cf$ can be expressed as the $n$th power of some $cf$.
In fact, using the characteristic function, the relation (B.1)
is transformed into
$$ [\hat p_\alpha (\kappa )]^n = \hat p_\alpha (c_n\, \kappa )\,
  \e^{\ds id_n\kappa}\,. \eqno(B.7)$$
The functional equation (B.7) can be solved completely and the solution
is known to be
$$ \hat p_\alpha (\kappa;\beta ,\gamma ,\delta )
 = {\exp} \l\{
 i\delta \kappa - \gamma^\alpha \, |\kappa |^\alpha \,
\l[ 1+i\, ({\sign}\, \kappa) \, \beta \, \omega (|\kappa |,\alpha)\r]
  \r\}	  \,,
\eqno(B.8) $$
where
$$ \omega (|\kappa |,\alpha) =
  \cases{
 {\tan}\, (\alpha\, \pi /2) \,,    &if $\q \alpha \ne 1\,, $\cr
 - (2/\pi) \, {\log}\, |\kappa|  \,, & if $\q \alpha =1\,. $\cr}
 \eqno(B.9)$$
Consequently {\it a random variable $X$ is said
to have a stable distribution if there are  four real parameters
$\alpha ,\beta,\gamma ,\delta $ with $0<\alpha \le 2\,, $
$-1\le \beta \le +1\,, $ $\gamma > 0\,, $  such that
its characteristic function has the  canonic form (B.8-9)}.
Then we  write
$p_\alpha (x;\beta ,\gamma ,\delta) \div
 \hat p_\alpha (\kappa;\beta ,\gamma ,\delta)$ and
$X \sim P_\alpha (x;\beta ,\gamma ,\delta )\,, $
so partly following the notation of Holt \& Crow (1973) and
 Samorodnitsky \& Taqqu (1994).
\vsp	
We note in (B.8-9) that  $\beta $ appears with different signs for
$\alpha \ne 1$ and $\alpha =1\,. $ This minor point has been the
source of great confusion in the literature, see Hall (1980) for
a discussion. The presence of the logarithm for $\alpha =1$
is the source of many difficulties,
 so this case  has often  to be treated separately.
\vsp
The $cf$ (B.8-9) 
turns out to be a useful tool for studying $\alpha$-stable
distributions  and for providing an interpretation of the additional
parameters,   $\beta $ ({\it skewness parameter}),
 $\gamma $ ({\it scale parameter}) and
 $\delta $ ({\it shift parameter}), see Samorodnitsky \& Taqqu (1994).
When $\alpha =2$  the $cf$ refers to
the Gaussian distribution with variance $\sigma ^2= 2\, \gamma ^2\,$
and mean $\mu =\delta \,;$  in this case
 the value of the skewness
parameter $\beta $   is not specified
because ${\tan}\, \pi =0\,, $ and
one conventionally takes $\beta =0\,. $
\vsp
One easily recognizes that a stable distribution is {\it symmetric}
if and only if $\beta =\delta  =0 $  and is symmetric about  $\delta $
if and only if $\beta =0\,. $
 Stable distributions with extremal values of the skewness parameter
 are called {\it extremal}.
One can prove
that all the extremal stable distributions with $0<\alpha  <1$
are one-sided, the support being $\RR_0^+$  if $\beta =-1\,, $
and  $\RR_0^-$	if $\beta =+1\,. $
\vsp
For the stable distributions $P_\alpha (x;\beta ,\gamma ,\delta )$
we now consider   the asymptotic behaviour
of the tail probabilities,
  $T^+(\lambda):= {\Prob}\,\{X> \lambda \}$ and
  $T^-(\lambda):= {\Prob}\,\{X<- \lambda \}\,, $
as $\lambda \to \infty\,. $
For the Gaussian case $\alpha =2$ the result is well known,
see \eg Feller (1957),
$$ \alpha =2\,: \q    T^\pm(\lambda )	\sim
   \rec{2\sqrt{\pi}\, \gamma }\,
   {\e^{\ds -\lambda^2/(4\gamma^2)}\over \lambda}\,,
 \q \lambda  \to \infty\,.
 \eqno(B.10)$$
Because of the above exponential decay all the moments of the
corresponding $pdf$ turn out to be finite, which is an exclusive
property of this stable distribution.
For all the other stable distributions
the singularity of the characteristic
function in the origin is responsible for the  algebraic
decay of the tail probabilities as indicated below, see \eg
 Samorodnitsky \& Taqqu (1994),
$$  0< \alpha <2\,: \q	  \lim_{\lambda \to \infty}\,
     \lambda^{\alpha}\,  T^\pm(\lambda )  =
   C_\alpha \, \gamma ^\alpha\, (1 \mp \beta)/2   \,,
 \eqno(B.11)$$
where
$$  C_\alpha = \l( {\ds\int_0^\infty \!\!\!
  x^{-\alpha }\, \sin x\, dx}\r)^{-1} =
   \cases{
  {\ds{1-\alpha\over \Gamma(2-a)\, \cos \,(\alpha \pi/2)}}\,,
      &if $ \alpha \ne 1\,, $\cr\cr
   2/\pi \,, &if $ \alpha =1\,. $\cr}\eqno(B.12)$$
We note that for extremal distributions ($\beta =\pm 1$)
the above algebraic decay holds true only for one
tail, the left one if $\beta =+1\,, $ the right one if $\beta =-1\,. $
The other tail is either identically zero if $0<\alpha <1$ (the
distribution is one-sided !), or exhibits an exponential decay
if $1\le \alpha <2\,. $
Because of the algebraic decay
we recognize that
$$     0<\alpha <2\,:\q
 \int _{|x|>\lambda } p_\alpha	(x;\beta,\gamma,\delta) \,dx
  = O(\lambda^{-\alpha })\,,\eqno(B.13)$$
so the	  absolute moments of a  stable non-Gaussian $pdf$
  turn out to be finite
if their order $\nu  $ is $0\le \nu   <\alpha $ and
infinite if $\nu   \ge \alpha \,. $
 We are now convinced that the Gaussian distribution is the unique
 stable distribution with finite variance.
Furthermore,  when $\alpha \le 1\,, $
the first absolute moment $ \langle {|X|} \rangle $ is
infinite  as well, so 
we need to use the  median to characterize the expected value.
\vsp
There is however a fundamental property shared by all the stable
distributions that we like to point out:
for any $\alpha $ the  stable  $pdf$ are $unimodal$ and indeed {\it
bell-shaped}, \ie
their $n$-th derivative has exactly $n$ zeros, see Gawronski (1964).
\vsp
We now come back to the $cf$  of a stable distribution,
in order to provide for $\alpha\ne 1$ and $\delta =0$
 a  simpler canonic form which allow us to derive convergent
and asymptotic power series for the corresponding $pdf$.
We first note that the two parameters $\gamma $ and $\delta $ in (B.8),
being related to a scale transformation and a translation,
 are not so essential since they do not change
the shape of distributions.  If we take $\gamma =1$ and $\delta =0\,, $
we obtain the so-called {\it standardized}  form  of the
stable distribution and   $X \sim P_\alpha (x;\beta ,1,0)$
is referred to as the $\alpha$-stable {\it standardized} random variable.
Furthermore, we  can choose the scale parameter $\gamma $ in
such a way to get from	(B.8-9) the simplified canonic form
used by Feller (1952, 1966-1971) and Takayasu (1990)
 for strictly stable distributions
($\delta =0$) with $\alpha \ne 1\,, $ which reads in
an  {\it ad hoc} notation,
$$  \hat q_\alpha  (\kappa; \theta) :=
  \int_{-\infty}^{+\infty}\!\! \e^{\ds i\kappa \, y}\,
  p_\alpha (y;\theta) \, dy
= \exp\, \l\{ {\ds - |\kappa |^\alpha} \, \e^{\ds\,\pm i\,\theta\,
     \pi/2}\r\}
 \,,\eqno(B.14)$$
where the symbol $\pm$ takes the sign of $\kappa \,. $
 This canonic form, that we refer to as  the
{\it Feller  canonic form}, is
derived from  (B.8-9)
if in addition to $\alpha  \ne 1$ and $\delta =0$ we require
$$ \gamma ^\alpha = \cos\,\l (\theta\,{\pi\over 2}\r)\,, \q
 {\tan} \l(\theta \, {\pi \over 2}\r) =
  \beta \, {\tan}\, \l(\alpha \, {\pi\over 2}\r)
\,. \eqno(B.15) $$
Here $\theta$ is the   {\it skewness}  parameter instead of $\beta $
and  its domain is restricted in the following
region (depending on $\alpha $)
$$ |\theta| \le \cases{
   \alpha \,, &if $\q 0<\alpha <1\,, $\cr
   2-\alpha \,, &if $\q 1<\alpha <2\,. $\cr}\eqno(B.16)$$
Thus, {\it when we use the Feller canonic form	for  strictly
stable distributions with index $\alpha \ne 1\, $  and skewness
$\theta \,, $
we implicitly select the scale parameter}
$\gamma$ ($0<\gamma\le 1$), which is related to
$\alpha\,, \,\beta$ and $\theta\, $ by (B.15).
Specifically,  the random variable  $Y\sim Q_\alpha(y;\theta)$
turns out to be related to the {\it standardized}  random variable
$X \sim P_\alpha (x;\beta ,1,0)$ by the following relations
$$ Y = X/\gamma \,,\qq
   p_\alpha (x;\beta,1,0) = \gamma \, q_\alpha (y=\gamma x;\theta) \,,
\eqno(B.17)$$
with
$$ \l\{
\begin{array}{lll}
\gamma &= {\ds [\cos\,(\theta \pi/2)]^{1/\alpha}}\,, \\ 
\theta &= {\ds (2/\pi)\,{\arctan}\,[\beta \,{\tan}\,(\alpha
	   \pi/2)]}\,, \\   
 \beta &= {\ds {{\tan}\, (\theta\pi/2) \over {\tan}\,(\alpha \pi/2)}}\,.
\end{array}\r.	\eqno(B.18)
$$
We recognize that $q_\alpha (y,\theta) =  q_\alpha (-y,-\theta)\,, $
so the {\it symmetric} stable distributions are
obtained if and only if  $\theta =0\,. $
We note that  for the {\it symmetric} stable
distributions  we get the identity  between
the {\it standardized} and the {\it L\'evy} canonic forms,
since  in (B.18)  $\beta =\theta =0\, $   implies $\gamma =1\,. $
  A particular but noteworthy case is provided by
$p_2(x;0,1,0) =  q_2(y;0)\,,  $
corresponding to the Gaussian distribution with variance $\sigma ^2=2\,. $
\vsp
The {\it extremal} stable distributions, corresponding
to $\beta =\pm 1\,, $
are now obtained  for  $\theta = \pm \alpha $  if $ 0<\alpha <1\,, $
and for $\theta =\mp (2-\alpha) $ if $1<\alpha <2\,; $
for them the scaling parameter turns out to be
$\gamma = [\cos\,(|\alpha|\,  \pi/2)]^{1/\alpha}\,. $
It may be an instructive exercise to carry out the inversion
of the Fourier transform  when $\alpha =1/2$  and
$\theta =-1/2\,. $
In this case we obtain the analytical expression for the
corresponding  extremal stable $pdf$,  known as the
(one-sided) {\it L\'evy-Smirnov} density,
$$ q_{1/2}(y; -1/2) =
{1\over 2\sqrt{\pi}}\, y^{-3/2}\, \e^{\ds - 1/(4y)}\,, \q y \ge 0\,.
\eqno(B.19) $$
The {\it standardized} form   for this distribution can be
easily obtained   from (B.19)
using (B.17-18) with $\alpha =1/2$ and $\theta =-1/2\,. $
We get
$ \gamma = [\cos\,(- \pi/4)]^{2} = 1/2\,, \;
   \beta  = -1 \,, $
so
$$ p_{1/2}(x;-1,1,0) = \rec{2}\, q_{1/2}(x/2; -1/2) =
 \rec{\sqrt{2\pi}}\, x^{-3/2} \,
    \e^{\ds - 1/(2x)}\,, \eqno(B.20)$$
where $x\ge 0\,,$
in agreement with Holt \& Crow (1973) [\S 2.13, p. 147].
\vsp
Feller (1952)
has obtained  from (B.14)
the following representations by convergent power
series for the stable distributions valid for $y>0\,, $
with
\par \noindent	$0<\alpha <1$ (negative powers),
 $$
q_\alpha (y;\theta) =
{1\over \pi\,y}\,  \sum_{n=1}^{\infty}
   (-y^{-\alpha})^n \, {\Gamma (n\alpha +1)\over n!}\,
  \sin \l[{ n\pi\over 2}(\theta -\alpha)\r]\,,
 \eqno(B.21)$$
\par\noindent
$1<\alpha \le 2$ (positive powers),
$$ q_\alpha (y;\theta )=
{1\over \pi\,y}\,  \sum_{n=1}^{\infty}
   (-y)^{n} \, {\Gamma (n/\alpha +1)\over n!}\,
  \sin \l[{ n\pi\over 2\alpha }(\theta -\alpha)\r]\,.
\eqno(B.22) $$
The values for $y<0$  can be obtained from (B.21-22) using the identity
$q_\alpha (-y;\theta )=  q_\alpha (y;-\theta )\,,\; y>0\,. $
As a consequence of the convergence in all of $\CC$ of the series
in (B.21-22) we recognize that
the restrictions
of the functions $y\,q_\alpha (y;\theta)$ on the two real semi-axis turn
out to be equal to certain {\it entire} functions of argument
$1/|y|^\alpha $ for   $0<\alpha   <1$ and
argument $|y|$ for  $1<\alpha \le 2\,. $
\vsp
It has	be shown, see \eg Bergstr\"om (1952), Chao Chung-Jeh (1953),
that the two series in
(B.21-22) provide also the asymptotic (divergent)
expansions to the stable $pdf$
with the  ranges of $\alpha $
interchanged from those of convergence.
 \vsp
From (B.21-22)	a relation    between stable $pdf$
with index $\alpha $ and $1/\alpha\,  $  can be derived
as noted in Feller (1966-1971).
Assuming $1/2<\alpha<1$ and $y>0\,, $  we obtain
$$ \rec{y^{\alpha +1}}\, q_{1/\alpha}(y^{-\alpha} ;\theta )
  =q_\alpha (y;\theta ^*)\,,  \q
  \theta ^*=\alpha(\theta +1)-1 \,. \eqno(B.23)$$
A quick check shows that  $\theta^*$ falls within the prescribed range,
$|\theta ^*|\le\alpha \,, $ provided that $|\theta |\le 2-1/\alpha \,. $
\vsp
We now consider two particular cases of the Feller series (B.21-22),
of particular interest for us,
which turn out to be related to the entire function of
Wright type, $M(z;\nu )$ with $ 0<\nu  <1\,, $
reported in Appendix A. 
These cases correspond to the following extremal distributions
  $$ \Phi_1(y) := q_\alpha (y;-\alpha) \,,
    \q y > 0\,,  \q 0<\alpha <1\,, \eqno(B.24)$$
$$ \Phi_2(y) := q_\alpha (y;\alpha-2) \,,
    \q y > 0\,,  \q 1<\alpha \le 2\,,  \eqno(B.25)$$
for which the Feller series (B.21-22) reduce to
$$
  \Phi_1(y) =
 {1\over \pi}\,  \sum_{n=1}^{\infty}
  (-1)^{n-1}\, y^{-\alpha n-1} \,
 {\Gamma (n\alpha +1)\over n!}\,   \sin \l({ n\pi\alpha}\r)\,,
  \q y > 0\,,
  \eqno(B.26)$$
 and
$$
 \Phi_2(y)=
{1\over \pi}\,	\sum_{n=1}^{\infty}   (-1)^{n-1}\,
   y^{n-1} \, {\Gamma (n/\alpha +1)\over n!}\,
  \sin \l({ n\pi\over \alpha }\r) \,, \q y > 0\,.
  \eqno(B.27)$$
\vsp
In fact, recalling the series representation of the general Wright
function,
$ W_{\lambda ,\mu }(z)$ with $\lambda > -1\,, \; \mu >0\,,$ see  (A.31),
and  the definition of the function $M(z;\nu)$ with
$0<\nu	<1\,, $  see (A.32-33),
we recognize that
$$ \Phi_1(y) =
     {1\over y} W_{-\alpha ,0} (-y^{-\alpha }) =
   {\alpha \over y^{\alpha +1}}\, M(y^{-\alpha};\alpha )\,,\q y > 0\,,
\eqno(B.28)$$
and
$$ \Phi_2(y)  = {1\over y}\, W_{-1/\alpha ,0} (-y)
 = {1\over \alpha }\,  M(y;1/\alpha )\,, \q y > 0\,.
 \eqno(B.29)$$
We would like to remark that the above relations with the Wright
functions have	been noted also by Engler (1997).
\vsp
 It is worth to point out  that,
whereas $\Phi_1(y)$ totally represents the
one-sided   stable $pdf$   $q_\alpha (y; -\alpha)\,,$
 $0<\alpha <1\,, $ with support in  $\RR_0^+\,, $
 $\Phi_2(y)$  is  the restriction
on the positive  axis of $q_\alpha (y; \alpha-2)\,,$
$1<\alpha \le 2\,, $ whose support is all of  $\RR\,.$
Since the function $M(z;\nu )$ turns out to be
 normalized in ${\RR}_0^+$, see (A.39-40),
we also note
$$ \int _0^{\infty} \Phi_1(y)\, dy = 1 \,;
 \qq
\int _0^{\infty} \Phi_2(y)\, dy = 1/\alpha \,.
\eqno (B.30)$$
\vsp
Using the results (A.41) and (A.37) we can easily evaluate the
Laplace transforms of  $\Phi_1(y)$ and $\Phi_2(y)\,, $ respectively.
We obtain
$$ {\cal{L}} [\Phi_1(y)] = \widetilde \Phi_1(s) =
  {\exp}\,(- s^\alpha )\,,  \q
    0<\alpha <1\,,
  \eqno(B.31)$$
$$ {\cal{L}} [\Phi_2(y)] = \widetilde \Phi_2(s) = {1\over \alpha }\,
 E_{1/\alpha }\,(- s)\,,
  \q	1<\alpha \le 2\,,   \eqno(B.32)$$
where $E_{1/\alpha }(\cdot)$ denotes the {\it Mittag-Leffler} function
of order $1/\alpha \,, $  see (A.23).
\vsp
It is an instructive exercise to derive the asymptotic	behaviours of
$\Phi_1(y)$ and $\Phi_2(y)$  as $y \to 0^+$ and $y \to +\infty\,. $
By   using the expressions $(B.28-29)$ in terms of the function $M$
and recalling the series and asymptotic representations of this function,
see (A.33) and (A.36), we obtain
$$ \Phi_1(y)  =
  \cases{
 O\, \l( y^{ -(2-\alpha )/[2(1-\alpha)]}\,
 {\e}^{\, - c_1\, y^{-\alpha /(1-\alpha)}}  \r) \,,
   &as $\q y\to 0^+\,, $ \cr\cr
  {\ds{\alpha \over \Gamma (1-\alpha)}}\, y^{\,-\alpha -1}\,
\l[ 1+ O\,\l(y^{ -\alpha }\r)\r]
   \,, & as $\q y\to +\infty\,, $\cr}
 \eqno(B.33)$$
$$ \Phi_2(y)  =
  \cases{
  {\ds{1/\alpha \over \Gamma (1-1/\alpha)}}\l[ 1+ O\,\l( y \r)\r] \,,
   &as $\q y\to 0^+\,, $ \cr\cr
  O\, \l( y^{ (2-\alpha )/[2(\alpha-1)]}\,
 {\e}^{ \,- c_2\, y^{\alpha /(\alpha-1)}}  \r) \,,
     & as $\q y\to +\infty\,, $\cr}
 \eqno(B.34)$$
where $c_1\,, \,c_2$ are positive constants depending on $\alpha\,. $
We note that the exponential decay is found for $\Phi_1(y)$
as $y\to 0^+$ but as $y\to + \infty$ for $\Phi_2(y)\,. $
 \vsp
Explicit expressions for {\it stable} $pdf$ can be derived
form   those   for   the
function  $M(z;\nu)$
when $\nu =1/2$ and $\nu=1/3\,, $ given in Appendix A,
see (A.34-35).
Of course the $\nu   =1/2$ expression can be used to recover
the well-known (symmetric) Gaussian distribution $q_2(y;0)$
accounting for (B.29), and the
(one-sided) L\'evy distribution  $q_{1/2} (y;-1/2)$, see (B.19),
accounting for (B.28).
The $\nu  =1/3$  expression provides, accounting for (B.28),
$$
  q_{1/3}(y;-1/3)  =
   3^{-1/3}\, y^{-4/3}\,
  {\Ai}\,\left[ (3y)^{-1/3}\right] 
    = {\ds  {1\over 3\pi}}\, y^{-3/2}\,
   {K}_{1/3}\,\l (2 / \sqrt{27y}\r),
 \eqno(B.35)$$
where $\Ai$ denotes the Airy function and
 $K_{1/3}$  the modified Bessel function
of the second kind    of order $1/3\,. $
The equivalence between the two expressions in (B.35) can be proved in
view of the  relation, see Abramowitz \& Stegun (1965-1972) [(10.4.14)],
  $$ {\Ai}\, (z) = {1\over  \pi} \sqrt{z\over 3}\,
  K_{1/3}\l( {2\over 3}\, z^{2/3}\r) \,.\eqno(B.36) $$
The case $\alpha =1/3$ has also been discussed by Zolotarev (1983-1986),
who has quoted the corresponding expression of the $pdf$
in terms of $K_{1/3}\,.$
\vsp
A  general  representation of all stable distributions (thus including
the {\it extremal} distributions above considered)  in terms of
special functions has been only recently achieved by Schneider (1986).
In his remarkable (but almost ignored) article, {Schneider}
has established that all the stable distributions  can be characterized
in  terms
of a general class of special functions,  the so-called  Fox $H$ functions,
so named after Charles Fox (1961).
For details on Fox $H$ functions, see \eg
the books Mathai \& Saxena (1978),  Srivastava \& Al. (1982)
 and the most recent paper by Kilbas and Saigo (1999). 
These functions are expressed in terms of special integrals in the complex-plane,
the {\it Mellin-Barnes integrals}\footnote{
The names refer to the two authors, who in the first 1910's
developed the theory of these integrals  using them
for a complete integration of the hypergeometric differential equation.
However, as pointed out  in  the  the Bateman Project Handbook
on High Transcendental Functions, 
see Erdelyi  (1953),
these integrals were first used
by S. Pincherle in 1888. For a revisited analysis of the pioneering work
of Pincherle (1853-1936, Professor of Mathematics at the
University of Bologna from 1880 to 1928) we refer
to the  paper by
Mainardi and Pagnini (2003).}.
